 \documentclass[preprint]{aastex}

\shorttitle{Bode, Ostriker, \& Turok}
\shortauthors{Warm Dark Matter}

\begin{document}


\title{Halo Formation in Warm Dark Matter Models}


\author{Paul Bode and Jeremiah P. Ostriker}
\affil{Princeton University Observatory, Princeton, NJ 08544-1001 }

\and

\author{Neil Turok}
\affil{DAMTP, Centre for Mathematical Sciences, Wilberforce Road,
Cambridge, CB3 0WA}


\begin{abstract}
Discrepancies have emerged between the predictions of standard
cold dark matter (CDM) theory and observations of clustering
on sub-galactic scales. Warm dark matter (WDM) is 
a simple modification of CDM in which the 
dark matter particles have initial velocities 
due either to their having decoupled as thermal relics, 
or having been formed via non-equilibrium 
decay. We investigate the nonlinear gravitational clustering
of WDM with a high resolution N-body code, and 
identify a number of distinctive observational
signatures. Relative to CDM, 
halo concentrations and core densities are lowered,
core radii are increased, and large halos emerge with 
far fewer low 
mass satellites. 
The number of small 
halos is suppressed, and those present are formed by 
`top down' fragmentation of caustics, as part of a `cosmic web' 
connecting massive halos. Few small halos
form
outside this web. 
If we identify small halos with dwarf galaxies, 
their number, spatial distribution,
and formation epoch appear
in better agreement with the observations for WDM than they are 
for CDM.
\end{abstract}




\section{Introduction}

Standard cold dark matter (CDM) theory appears to be in conflict with 
observations of clustering on sub-galactic scales. First, 
the theory predicts  a very large number of
low mass halos.
For example, of order 500 satellites of mass greater than 
$10^8 M_\odot$ are expected within a halo as massive as that
of the Milky Way
whereas only 11 candidate dwarf galaxies with comparable mass are observed
\citep{Moore}.
In the whole Local group, 
out to $\sim 1.5$ Mpc there are only $36$ known dwarf galaxies
\citep{mateo}, where several thousand would be expected 
in a straightforward interpretation of the theory.
The second problem with CDM is that simulations produce
halos with more concentrated cores than those inferred 
from the rotation curves of real galaxies
\citep{Moorecore,Ghigna99}.
There are a number of related but more complex problems.
For example, the predicted satellites would lead to a thickening
of the disks of galaxies which is not observed. And the dense
cores,  which form very early, would cause baryonic cooling
to be too efficient, 
leading to disks 
with an order of magnitude too little angular momentum 
\citep{Dolg,Navarro}.

These problems may well have complex astrophysical
solutions.
Feedback processes such as heating and supernovae winds 
might inhibit star formation in low mass halos \citep{bullock,silk},
although it is not clear why such processes would suppress the
number of satellites without destroying them all.
No convincing mechanism has been devised to explain the fact
that most observed dwarf systems have characteristic densities
which are considerably lower than
those of dark matter halos of the same mass
predicted from $\Lambda$CDM theory, and it certainly 
seems implausible that
feedback processes would preferentially inhibit star formation
in the denser halos. Further, dwarf spheroidals
are believed to be completely dominated by dark matter,
but their stellar profiles
show no sign of a central density cusp \citep{irwin}.

The observational situation regarding 
galaxy cores has been rapidly evolving.
It was recently claimed that beam smearing effects were significant,
bringing observations of the inner rotation curves of low surface 
brightness galaxies 
closer to the predictions of CDM \citep{vdbswat00,swaters} 
(although there remained a discrepancy with the 
$\rho \propto r^{-1.5}$ central profile indicated  by
high resolution CDM simulations \citep{Moorecore,swaters,klypin}).
But very recently de Blok {\it et al.} \citep{deblok} 
reanalised the same data and added considerably more new data, only
to reach the
opposite conclusion that beam smearing effects were not important
and the data are seriously
discrepant even with
the milder $\rho \propto r^{-1}$ behavior of the NFW profile
\citep{NFW97}.

There is another potential problem with CDM, noted some time ago
by \citet{pjep00,peeb01}.
The puzzle is that faint dwarf
galaxies as measured in our vicinity (for a picture of 
the local volume see \citet{peeb} p. 26 and \citet{SRE97})
do not fill the voids in the bright
galaxy distribution. Rather they tend to follow an even sharper 
distribution than the bright galaxies, tracing out surprisingly thin sheets
around the voids. This effect has not as far as we know been systematically
quantified, but is interesting in connection with the satellite 
problem of CDM. That is, observationally, 
dwarf galaxies seem both more strongly correlated with 
bright galaxies than in CDM, whilst at the same time being
suppressed within large galaxy halos. 
In CDM theory both features are hard to understand, since 
the smaller halos formed earlier and should not 
be strongly correlated 
with later forming, larger scale structure. They
should fill both the voids and massive structures alike, as we shall verify 
in our $\Lambda$CDM simulations
presented below.  Even if supernova winds or heating processes
inhibit star formation in dwarf halos, and thereby suppress
dwarf galaxies, it is 
hard to see why these
processes would act preferentially in the voids.
Another clue of potential significance is that 
observations \citep{metc} point to the formation of dwarf galaxies
at $3<Z<4$, later than bright galaxies.  This is not what one
expects in a hierarchical
theory like CDM where small objects should form first. 

If these problems with CDM are real,
they represent a remarkable opportunity. The observed pattern
of gravitational clustering may be revealing the physical properties
of the dark matter. If so, this 
will be an invaluable clue to
physics beyond the standard model. Deciphering this clue 
represents an exciting challenge in which both 
more refined observations and numerical simulations will be needed.
Even if the conclusion of this work is that CDM theory survives,
it will have been strengthened in the process. Alternative
theories are valuable foils against which the successes of the
better theory may be judged. 

The small scale problems mentioned above
do not negate
the remarkable successes of the
$\Omega_{M}\sim 0.3$, 
$\Omega_{\Lambda} \sim 0.7$ $\Lambda$CDM cosmology
on larger scales \citep{BOPS99}. These include the
abundances 
and observed evolutionary properties
of Lyman $\alpha$  clouds, large galaxies, clusters and of course
the normalization and shape of the 
power spectrum of cosmic microwave anisotropies. 
Therefore it seems sensible to seek a modest
modification of the scenario whose sole effect is to damp small scale
structure. 

Free streaming due to thermal motion of particles,
or Landau damping, is the simplest known mechanism for smearing out 
small scale structure.
This process was central to the
hot dark matter (massive neutrino) scenario for structure formation,
which
failed because the particle speeds were too high and 
erased perturbations up to tens of Megaparsec scales. 
Structure formation was `top down', with galaxies forming only 
through fragmentation of pancakes, and at
redshifts too low to be 
compatible with observation.

Warm dark matter (WDM) is just hot dark matter cooled down. 
We shall review below some of the many ideas in the literature as to how
WDM might be produced. A particularly interesting possibility is that 
if the reheating temperature following 
inflation is low, then
standard $\mu$ or $\tau$ neutrinos can be warm dark matter.
Historically, WDM  was
considered (and quickly rejected)
as a means of reconciling computed and observed
cluster abundances while maintaining
a critical density for the dark matter 
\citep{widrow}. That problem is resolved in the $\Lambda$CDM model
by simply lowering the
dark matter density. But the new problems faced by CDM
motivate reconsideration of WDM 
(reincarnated now as $\Lambda$WDM) with a view to 
damping structure on sub-Megaparsec  scales \citep{Dolg,hogdal}, requiring much
lower particle velocities.

It is conventional to discuss WDM in terms of 
a canonical candidate, a light fermion with two spin degrees of
freedom like the neutrino, assumed to 
decouple adiabatically and whilst still relativistic. 
The number of relativistic degrees of freedom at decoupling 
determines the effective temperature
of the warm particles today,  and the dark matter density
determines the WDM particle mass $m_X$.
We shall discuss WDM
in terms of this mass,
but it should be remembered that 
what really matters
is the particle streaming speed,
given for the canonical particle in equation (\ref{eq:veq}) of the Appendix.

We discuss linear perturbation theory relevant to warm particles
in the Appendix, where we derive the following formula 
for the smoothing scale, defined as the comoving half-wavelength 
of the mode for which
the linear perturbation amplitude is suppressed by two:
\begin{equation}
R_S\approx 0.31\,
\left({\Omega_X\over.3}\right)^{.15} \left({h\over .65}\right)^{1.3} 
\left({{\rm keV}\over m_X}\right)^{1.15} h^{-1} {\rm Mpc}.
\label{eq:rsup}
\end{equation}
The scaling given here is different from that given in the previous literature
on WDM (which mostly refers to \citet{bs} and \citet{bbks}).
The scaling 
usually given does not 
properly take into account streaming 
prior to matter-radiation equality.

The outline of this paper is as follows. In Section 2 we review 
particle physics mechanisms through which WDM
may be formed.
In Section 3, we discuss the lower bound imposed on $m_X$
by the requirement of early structure formation
and in particular of reionizing the universe (the Gunn-Peterson constraint). 
In Section 4 we discuss the phase space density (Tremaine-Gunn)
constraint, and show that the initial thermal velocities of the
particles are only relevant to the inner cores of halos, scales of
hundreds of parsecs.
The following sections are devoted to 
numerical results, revealing 
several potential observational 
signatures of warm dark matter.

We find that replacing cold with warm dark matter has the following 
effects:

\begin{enumerate}
\item    Smoothing of massive halo cores, lowering
core densities and increasing core radii.
\item    Lowering greatly the characteristic density of low mass halos.
\item    Reduction of the overall number
of low mass halos.
\item    Suppression of the number of low mass
satellite halos in high mass halos.
\item    Formation of low mass halos almost solely within caustic 
pancakes or ribbons connecting larger halos in a `cosmic web'. 
Voids in this web
are almost empty of small halos, in 
contrast to the situation in CDM theory.
\item    Late formation ($Z<4$) of
low mass halos, in a `top down' process.
\item    Suppression of halo formation at high redshift ($Z>5$), and
increased evolution of halos at lower redshifts relative to
CDM.
\end{enumerate}

The first four findings indicate that WDM  holds
some promise as a solution to the satellite and core density 
problems of CDM. The fifth is
interesting as it may solve the  problem raised
by Peebles. The sixth and seventh items point to observational tests at 
high redshift, where there is the possibility of actually seeing
the pancake formation as it first occurs. Recent HST observations
probing galaxies at high redshift seem to  indicate the 
formation of dwarf galaxies after bright galaxies \citep{metc}, which 
is an argument in favor of WDM.
It is perhaps worth emphasising that 
{\it WDM succeeds both in 
suppressing the number of
satellites in large halos, whilst producing a distribution
of dwarf halos in a cosmic web connecting 
higher mass galaxies.}
All of these effects need to be 
better quantified in both 
observational data and simulations, so that 
precise statistical tests are possible. 

It is a challenging problem at the limit of current numerical
codes to simultaneously  represent the modes responsible for structure
formation (tens of Megaparsecs at least) whilst resolving 
objects as small as dwarf galaxies
$< 10^8 M_{\odot}$. We chose the compromise of studying 
large boxes of $20 h^{-1}$ Mpc with $256^3$ particles in which
we compare $\Lambda$CDM with $\Lambda$WDM, with 
WDM particle
masses of
$m_X=$ 175eV and 350 eV corresponding to 
smoothing scales $R_S$ of 2.3 and 1.0 $h^{-1}$ Mpc respectively.
These values for $m_X$ are almost certainly excluded by the
Gunn-Peterson limit which we detail below, or by 
the limit from Lyman $\alpha$  clouds, pointing to
$m_X>$ 750 eV \citep{MaSper}. 
Nevertheless they are helpful for illustrating the qualitative
differences 
between WDM and CDM clustering on large scales. 
We also performed a simulation with a more 
realistic value of $m_X=$ 1.5 keV, 
corresponding to a smoothing scale of 0.19
$h^{-1}$ Mpc, in smaller 
$3 h^{-1}$ Mpc boxes with 
$128^3$ particles. Finite size effects are important in this box,
since long wavelength modes in the power spectrum are not
properly represented. Nevertheless the simulation shows 
that the effects found at low $m_X$ persist at higher values.
A second limitation is that with the current code,
computation time becomes prohibitive for the $256^3$ boxes
at redshifts below  unity. Therefore our
largest simulations end at $Z=1$.
The signatures we identify deserve
more detailed study, and larger simulations  will be carried out
in future work.\footnote{
Movies of the simulations were found to be very instructive in showing
how structure formation in WDM proceeds.
We have made
these, higher resolution versions of the figures and some 
other results  not included here available on the web, at 
\url{http://\allowbreak
astro.\allowbreak
Princeton.\allowbreak
EDU/\allowbreak
$\sim$bode/\allowbreak
WDM/}.
}

Simulations have also been recently carried out by
\citet{ACVDF01}, 
\citet{CoAvVa00}, and \citet{colin}.
They found that WDM results in a reduction of the number of satellites
in Milky-Way sized galactic halos, and in satellite halos
with less centrally concentrated density profiles.
The work presented here is in agreement with these findings.
\citet{Moorecore} simulated a $\sim 10^{12} M_\odot$
halo and found little difference in the density profile
between warm and cold dark matter. \citet{WhCr00}
used simulations to examine the nonlinear power spectrum in
WDM models.

\section{Warm Dark Matter}

In WDM, 
free streaming 
smears out perturbations on scales smaller than the comoving scale $\Delta x$ 
over which
a typical particle has traveled. In a flat FRW universe, the momentum
$p$ of a particle
scales inversely with the scale factor $a$. This may be understood
as a stretching of the de Broglie wavelength with $a$. The peculiar
speed is given by the usual formula $v_p = p/\sqrt{p^2+m_X^2}$, 
with $p= m_X a_{nr}/a$, where
$a_{nr}$ is the scale factor when the particle
becomes non-relativistic (i.e. when $v_p=c/\sqrt{2})$.
The motion in comoving coordinates
is given by $d x/ dt = v_p/a \approx c a_{nr}
a^{-2}$ at late times. In the matter era, 
where $a \propto t^{2\over 3}$,
the comoving displacement $\Delta x$
converges. A crude estimate of the smoothing length is
obtained by multiplying the 
speed of the particle at matter-radiation equality
by the comoving horizon scale at that time. Up to a 
numerical factor this is the smoothing scale usually quoted 
\citep{bs,bbks};
\begin{equation}
R_{>eq} \approx 0.2\, (\Omega_X h^2)^{1\over 3} 
\left({1.5\over  g_X}\right)^{1\over 3}
\left({{\rm keV} \over m_X}\right)^{4\over 3} {\rm Mpc}.
\label{eq:geq}
\end{equation}
However the situation is 
actually more complicated since for the warm particle speeds
of interest
the particles become nonrelativistic long before 
matter-radiation equality. There is thus a substantial period of time
in the radiation era
during which the Jeans mass is low and linear perturbations grow
logarithmically with time, whilst the comoving displacement 
is also growing logarithmically (since $a\propto t^{1\over 2}$ in the
radiation era).
The linear evolution of perturbations during this epoch may be 
understood using the nonrelativistic Gilbert equation 
discussed in the Appendix, which 
leads to the improved scaling formula given in (\ref{eq:rsup}) above,
and which we used in the simulations described below. But
for a crude discussion, as we give now, the 
the scaling of the smoothing length
(\ref{eq:geq}) is certainly adequate. 

A review of the mechanisms through which WDM may be formed
and their relation to cosmology can be found in 
\citet{widrow}. 
Warm dark matter is most simply discussed in comparison with 
the usual hot dark matter scenario.
In the hot dark matter scenario (with a single massive neutrino)
the particle mass would be 
$12$ eV. These neutrinos would have thermal speeds of 
$v_s/c\sim 0.2$ at equality, and their motion would erase 
structure on comoving scales of 
$\sim $ 25 Mpc. If we wish to preserve the successes of CDM theory
on comoving scales greater than say 0.1 Mpc, we need to reduce the streaming
speed by a factor of 250.
(In this discussion we shall choose cosmological 
parameters $\Omega_{DM}=0.3$ and $h=0.65$, unless otherwise stated.) 

For a thermal relic particle which is 
relativistic at decoupling but nonrelativistic by equality, 
the typical streaming speed at equality is 
$v_s/c \propto
T_X/m_X$, where $T_X$ is an effective temperature varying 
inversely with the scale factor. Neglecting the subtleties of
logarithmic growth in the radiation era, this speed determines the
streaming length.
To lower $v_s$ by 250 we need to either lower $T_X$ or raise $m_X$. 
However, if the particle decouples with thermal
abundance, then we have $\rho_X \propto T_X^3 m_X$ which
must be held fixed to
explain the dark matter today. Thus to obtain the desired reduction
in $v_s$ we must 
lower $T_X$ by $\sim$ 4 and increase 
$m_X$ by 
$\sim $ $4^3=64$, bringing it close to 1 keV.

The factor of $\sim$ 4  reduction in $T_X$ 
could be produced in several ways. The simplest is that
the $X$ particles decoupled when there were 64 times
the number of relativistic degrees of freedom than when neutrinos decoupled,
i.e. $64\times 10.75=688$. This requires several times more 
degrees of freedom than present in the
standard model (106.75) or supersymmetric versions ($\sim 200$). A
substantially larger number is certainly possible in 
theories with larger gauge groups and representations,
especially those with extra dimensions. As far as we know, no
detailed computations of entropy release have been performed in the latter 
case,
but at temperatures above 1 TeV the large number of Kaluza-Klein modes
would surely become important.
If the dark matter particle possesses only gravitational
strength couplings and therefore decouples early, it would be
left with a substantially lower effective temperature than the radiation
fluid. 
A plausible candidate
would be the gravitino \citep{kawasakio}.

In fact, all that is really required is that there be some
release of entropy after $X$ decoupling. 
One possibility is that 
the universe became
dominated by some form of
matter with $P< {1\over 3} \rho$ after $X$ decoupling.
For example, a mild first order 
phase transition with some supercooling
and latent heat release, a rolling scalar field which 
dominated briefly before being converted to radiation, or 
a massive species which dominated before decaying
into radiation but not $X$'s. This must happen when the temperature 
was greater than 
$\sim$1 MeV if it is not to disturb nucleosynthesis.

Another apparently viable scenario, discussed recently 
\citep{kawasaki,gkr}
is that the reheating temperature after inflation $T_{RH}$ was 
low, so that the maximum temperature of the radiation 
era was in fact only a few MeV. In this case, if the neutrinos 
are only produced via electron positron annihilation, 
their number density can be considerably lower than in the
usual equilibrium scenario. As \citet{kawasaki} point out,
although this reduces the expansion rate of the universe which
decreases the helium abundance,
there is a compensating
effect because there are fewer neutrinos to convert 
neutrons to protons. According to \citet{gkr}, in such a
scenario, ordinary neutrinos
may be resurrected as warm dark matter. \citet{gkr} give 
$\Omega_\nu h^2 = (m_\nu /{\rm 70 \,keV}) (T_{RH}/{\rm MeV})^3$ for the
density of $\mu$ and $\tau$ neutrinos
where $m_{\nu}$ is the neutrino mass.
(Unfortunately \citet{kawasaki} and \citet{gkr} are apparently 
inconsistent in their neutrino densities by over an order of magnitude. 
We understand the origin of the discrepancy is still unclear 
\citep{giuprivate}).
Recalling that the desired momentum 
(in units where $c=1$)
is $\sim 10^{-3} m_\nu$ at equal density, it is
$\sim 10^3 m_\nu T_{RH}/{\rm MeV}$ when the neutrinos are produced.
The neutrinos are
radiated from electron-positron and neutrino-antineutrino pairs in the hot 
plasma
with thermal momenta $p \sim 3 T_{RH}$ assuming $T_{RH}>> m_\nu$.
These equations 
imply 
$m_\nu \sim 3$ keV, and for
$\Omega_\nu \sim .3$ and $h \sim 0.65$ we obtain
$T_{RH} \sim 2$ MeV. The numerical coincidence here is
very interesting, and deserves further study. Note also that
in this scenario, the number densities of $\mu$ and $\tau$ neutrinos are 
suppressed, so that matter-radiation equality is now
at a higher redshift (by about thirty per cent). 
This is helpful in 
that there is more time for structure to form, alleviating the
reionization bound discussed below. Interestingly, it  also shifts 
the  Doppler peak of the CMB anisotropy to
lower $l$.

A different alternative is that 
$X$ particles carry a conserved charge like baryon number, and 
an $X$ asymmetry is formed as a result of 
non-equilibrium $CP$ violating processes.
In this case we would have (as for baryons) $\rho_X$ 
$\sim \epsilon m_X T_X^3$ with
$\epsilon$ a dimensionless
$CP$ violating coupling. Assuming the dark matter particles
have the same temperature as neutrinos, slowing them down by 
the required amount would require their mass be increased by
250, thus we would have $m_X
\sim 3$ keV, and the 
required dark matter density is attained for 
$\epsilon \sim .004$. 

One may also consider $X$ particles produced via 
entirely non-thermal mechanisms. One idea
\citep{Thomas} is that a supersymmetric condensate simultaneously decays
into 
the lightest super-particles, which form the dark matter,  and baryons,
in a ratio which 
explains the observed coincidence in their densities today.
If the  condensate mass were of order $300$ times
the dark matter particle mass, and if the decay took place at 
around 1 MeV, then the dark matter particles would have gamma
factors of  
$p/m_X \sim 300$ at that time, and speeds of $
3 \times 10^{-4} c$ at matter-radiation equality, 
equivalent to a 1 keV 
warm particle.

Other examples of non equilibrium production
of warm particles are resonant production
of sterile neutrinos via oscillation from active neutrinos
\citep{Dolgn,AbazFP01}, and particle production from the decay
of massive particles or cosmic defects \citep{Brandenberger}.

Even from this limited discussion it is clear that
there is an (over) abundance of mechanisms for producing warm
dark matter, using modest extensions of physics
beyond the standard model. 

\section{Early Structure Formation in WDM}

If one uses WDM to solve CDM's small scale problems
there is a window of allowed streaming speeds: too
low and the WDM is just like CDM;
too high and structure formation begins too late, as in HDM.
A particularly stringent limit is the Gunn-Peterson bound
\citep{GP}, provided by the fact
that quasar spectra reveal little absorption from 
smoothly distributed atomic hydrogen.
Thus the intervening regions of the 
universe must have been almost completely ionized.
The highest redshift quasars (at $Z=5.5$ and $5.8$) 
have only recently  been discovered
\citep{Stern,Fan}, and have normal spectra. There must therefore have been
enough UV radiation released from stars or
AGN
to fully ionize the universe by this epoch \citep{GP}.

The fraction of the baryons which could plausibly have undergone
gravitational collapse and star formation by a given redshift is 
easily computed (assuming the primordial fluctuations 
are Gaussian) from the power spectrum given in the Appendix. 
The fraction of the universe $f_{nl}$ attaining an
overdensity greater than $\delta_c$ at some redshift is just
probability that $\delta>\delta_c$ at any one point. We choose
$\delta_c$ to be the linear theory amplitude corresponding to
collapse to zero radius 
in the nonlinear spherical collapse 
model. 
In an 
$\Omega_M=0.3$ cosmology, one has $\delta_c \approx 1.675$ \citep{ECF96}, 
slightly smaller than the canonical critical density value of 1.69.
We compute the 
fraction $f_{nl}$ by integrating the power spectrum and employing
the linear theory growth factor appropriate to this cosmology. 
For $Z>2$ the growth factor $g(Z) \approx (\delta_0 (1+Z))^{-1}$
where $\delta_0$ is the loss of growth relative to an $\Omega_M=1$
flat universe, at $Z=0$. The latter is well approximated \citep{lahav} by
\begin{equation}
\delta_0 \approx 2.5 \Omega_M\left[\Omega_M^{4\over 7} -\Omega_\Lambda
+(1+{1\over 2} \Omega_M)(1+{1\over 70}\Omega_\Lambda)\right]^{-1}.
\label{eq:lahav}
\end{equation}
The result for $f_{nl}$ at redshifts $7,6,5,4,3$ is plotted
as the solid curves in Figure (\ref{fig:ion})
as a function of
the warm particle mass,  assuming the canonical neutrino-like particle
and a $\Lambda$WDM theory normalized to COBE ($\sigma_8=0.9$).

When baryons are bound into stars, 
nuclear burning releases on order $4 $ MeV per baryon. To ionize
hydrogen, $13.6$ eV is needed, thus a conservative bound
is that at least $3.4 \times 10^{-6}$ of the universe must have undergone
gravitational collapse in order to re-ionize the universe.
(Even this limit is not completely 
watertight, since
black holes can release more energy per gram than stars).
A more realistic bound is obtained
by assuming that the first generation of stars was typical
of star forming regions today: for such a population 
nuclear burning
releases on order 4000 ionizing photons per baryon \citep{Madau}.
This leads to a bound  $f_{nl} > 2.5 \times 10^{-4}$.
A greater nonlinear 
fraction is needed if one allows for recombination in denser clumps.
Gnedin and Ostriker \citep{GO97}
find (for $\Lambda$CDM) that 10 photons
above 13.6eV are required for each net ionization.
Note however
that a WDM universe is very smooth at early times 
between the high peaks, so
clumping is likely to be less significant than
in CDM \citep{Madau}.

The fraction $f_{nl}$ is very sensitive
to the value of $\sigma_8$ since the latter occurs squared 
in the exponent of the Gaussian probability distribution.
In Figure 1 we show the results of increasing
$\sigma_8$ from $0.9$ (the result for a COBE-normalized $n=1$ 
primordial spectrum) to $\sigma_8=1.0$. The latter
would be appropriate for COBE
normalization with
a small positive tilt, $\delta n \sim 0.02$. At present such a tilt cannot
be ruled out, although future CMB data may eventually allow one to do so.
Figure 1 shows that  a minimal warm
mass of order 400 eV (700 eV) is needed for $\sigma_8=1.0$ (0.9),
if one allows an additional factor of 3 for recombination.
A more detailed treatment of the reionization
limit leads to the requirement $m_X \gtrsim 0.75$keV \citep{BHO01}.

The way in which structure forms in WDM is conveniently illustrated
in Figure 2. This shows the 
rms mass fluctuation $\sigma_M$ in linear theory, computed in a 
spherical top hat window of varying radius, plotted against the 
comoving mass contained in the window.
It is given in units of the 
linear theory growth factor 
$g(Z)$, normalized to unity at $Z=0$.  For the cosmology of interest,
and for $Z>2$, $g(Z)$ is well approximated by
$\left(\delta_0(1+Z)\right)^{-1}$, where
$\delta_0=0.778$.

Above the WDM smoothing scale the theory behaves
like CDM. But as the mass scale is lowered, the rms mass fluctuation tends
to a constant, just because the density
field is smooth on small scales. Gravitational collapse 
occurs in most of the mass in objects whose scale is 
of order the WDM smoothing scale. These objects collapse 
when the linear theory rms amplitude $\sigma_M$ reaches of order unity.
From Figure 2 and the values of $g(Z)$ given above 
we see that the corresponding redshift 
increases roughly from 3 to 5 to 7
as the 
warm particle mass is increased from 0.5 to 1.5 to 3 keV. 
So in WDM one expects to see a great deal more evolution of 
galaxy scale objects at these redshifts than in CDM.

What about the formation of dense collapsed objects like the centers
of bright galaxies? In the simplest treatment, using the spherical
collapse model, these regions have an overdensity of
$\sim 200$ at the moment of collapse and thereafter their overdensity
scales as $(1+Z)^3$. Thus halos collapsing 
at redshifts $\sim 4$ could plausibly attain overdensities of
a few times $10^4$ today, and with the inclusion of mergers
 this is within reach of
the observed overdensities of the inner parts 
($r<10 h^{-1}$ kpc) of bright galaxies as inferred from
rotation curves ($v_{rot} \sim 200$ km \,s$^{-1}$ implies an overdensity 
of $\sim 3 \times 10^5$). 
Since these regions contain a mass of $\sim 10^{11} h^{-1}
M_{\odot}$, and the number density of such galaxies is $\sim 10^{-2}$ 
$h^3$ Mpc$^{-3}$, for $\Omega_M \sim 0.3$ they constitute around 1 per cent
of the mass in the universe. 
From Figure 1, we see that 
for warm particle masses greater than 700 eV, and with $\sigma_8=0.9$, 
more than one per cent of the
dark matter collapsed by redshift 4. But as we lower the warm particle mass, 
the curves steepen and the corresponding redshift falls
sharply. Thus it is clear that warm particle masses lower than
$\sim 500$eV are strongly constrained by the requirement that dark matter
halos with the observed properties are formed.

Now consider  smaller objects. Dwarf galaxies, like Draco and Ursa Minor
have half masses of $\sim 1.0 \times 10^7 M_{\odot}$ and half mass  radii
of $\sim 500$pc \citep{svdbbook}.
This corresponds to a characteristic
density of $2 \times 10^{16} M_{\odot}/{\rm Mpc}^3$ and an
overdensity of $5 \times 10^5$ for currently favored cosmologies;
they are believed to be dominated by dark matter.
They also have very
low line of sight
velocity dispersions, $\sigma \sim 9 $ km \,s$^{-1}$ \citep{mateo}.
Small objects like this are only formed in WDM through fragmentation
of caustics at late times in the vicinity of larger mass galaxies. 
As we discuss in the next section, their 
densities are 
perfectly compatible with the Gunn-Tremaine bound (which yields
a mass limit $m_X> 400$eV for the canonical warm candidate).
But how can such cold, relatively dense objects 
ever form? The most plausible location is at 
the center of caustics. For cold planar collapse,
the density of the virialized 
sheet diverges as the normal distance $z$ tends to zero, as $z^{-{1\over 2}}$
\citep{peeb}.
If a particle turns around at coordinate $z$
at time time $t_c$, then its peculiar velocity is $\sim z/t_c$.
If the turnaround time is at $Z=4$,
$t_c \sim 0.11 H_0^{-1}$ for $\Lambda$WDM with $\Omega_M=0.3$.
Then if $z$ is the comoving scale corresponding to the mass
of a dwarf, $d \sim 0.07/(1+Z) \sim .01$ Mpc, and we find a 
collapse velocity of $\sim 6.5$ km\, s$^{-1}$, similar  to
what is observed. Clearly this issue requires further study,
but the estimate is not discouraging. 

It is interesting to compare inferred dark halo overdensities over the
full observational range, from
dwarf galaxies to rich clusters of galaxies. As an example of a 
well studied cluster, see \citet{rines}. The mass is $\sim 10^{15}
h^{-1} M_\odot$, and the inferred half-mass overdensity is
$2.5\times 10^3$ in this cosmology. (Here we are being conservative,
in including the entire infall mass outside the virial radius.)
The usual
scaling argument applied to hierarchical structure formation
(e.g. \citet{peeb}, p. 628) predicts
that the overdensity for  collapsed halos 
$\rho_{halo} \propto M_{halo}^{-(n+3)/2}$
where $n$ is the effective power law for the linear theory power
spectrum ($P_k\propto k^n$) on the relevant scale.
The rms mass fluctuation $\sigma_M
\propto M_{halo}^{-(n+3)/6}$. Comparing $\sigma_M$ on 
the scale of clusters $\sim 1.5 \times 10^{15} M_\odot$ and dwarf galaxies
$\sim 2\times 10^7  M_\odot$, for $\Lambda$ CDM one obtains an effective index 
$n\sim -2.2$. The theory predicts a ratio of $\sim 1400$ for the 
characteristic dark matter densities of clusters and dwarfs of these
scales. 
The observed
ratio is 200, which is considerably 
smaller.
Clearly a larger data base
would be desirable, but this comparison points to
a possible further problem with CDM. 
 
\section{The Phase Space Constraint}

Liouville's theorem, stating that the phase space density is conserved
along particle trajectories in a collisionless fluid, provides a
fundamental constraint on the clustering of warm particles
\citep{GT}, which we review below. \citet{hogdal} 
and \citet{Sellwood00}
have recently discussed the
relevance of this constraint to WDM, which unlike CDM has a finite
primordial phase space density due to the thermal particle velocities.
However, by the time of halo 
collapse the thermal particle speeds are small and 
have little influence on halo formation 
except 
in the very inner-most regions of massive halos.
The smoothing of the
primordial power spectrum which occurred at very early times,
and which may
be treated in linear theory as we do in the Appendix, is the main effect.

Most of the mass undergoes gravitational collapse 
at modest
redshifts in WDM, $Z<5$. By this time, the warm particle speeds
are very low. From  
(\ref{eq:veq}) in the Appendix,
the rms speed for a 1 keV particle at $Z=5$ 
would be only $v=$ 0.25 km\, s$^{-1}$, 
far lower than the velocities induced in gravitational collapse in
any of the objects we are interested in today. Thus heating due to 
gravitational infall vastly outweighs the initial thermal energy,
even allowing for the increase due to phase space density
conservation ($v \propto \rho^{1 \over 3}$).

What little remains of the primordial velocities could really only
affect the far interiors
of massive halo cores. To estimate this effect, consider
the infall of a warm particle falling from initial radius
$r_i$ into a pre-existing halo.
The principal effect of the thermal motion is to give the particle
angular momentum, producing a centrifugal barrier
keeping the particle away from $r=0$;
only for radii inside this barrier is the
structure of the halo significantly altered relative to a pure CDM halo. 
This barrier is only significant for radius $r$ smaller than 
$r_{min} \approx r_i v_\perp/v_{max}$, where $v_{max}$ is the 
maximal radial speed induced by falling through the halo and
$<v_\perp^2>= {2\over 3} <v^2>\equiv {2\over 3} v_{rms}^2$ 
is the initial thermal transverse velocity dispersion.
Assuming for simplicity a halo with a flat rotation curve and
density run $\rho = \rho_s (r_s/r)^2$, 
$v_{max} \approx \sqrt{8\pi G \rho_s r_s^2 \,{\rm ln}(e r_i/r_{min})}$. 
We should take the initial radius $r_i$
to be a little larger than the radius of the shell now virialized at $r_s$ 
when it turned around at time $t$, so $r_i \sim  2 r_s$. 
If the halo has just virialized,
the mean overdensity in $r_s$ is $\sim 150$, which implies that
$\rho_s \sim 50 \rho_b$ where the background matter density 
$\rho_b= 1/(6\pi G t^2)$.
Thus we find 
$r_{min} \sim f v_{rms} t$ where $t$ is the time when the particle
turned around and $f$ is of order 0.1 divided
by the square root of the logarithm. 
For collapse at $Z=5$ ($t=.09 H_0^{-1}$) this 
gives a scale $r_{min} = 250 f \sim 25 h^{-1}$ pc, smaller
than scales of interest.
Generalizing this calculation to other halo profiles is straightforward. For 
example,
assuming an NFW profile (\ref{eq:NFW}) and 
$r_{min}<<r_s$, one finds $v_{max} \approx \sqrt{32 \pi G \rho(r_s) r_s^2}$,
so the conclusions are very similar.

If we
considered infall into a less symmetrical halo, we would still obtain
$f < 1$. Thus it seems safe to conclude that for warm particle
masses larger than $1$ keV, thermal velocities are
unimportant in determining the structure of halos on scales of a kiloparsec
or above. They may be relevant 
to the formation of very small halos - if
$f \sim 1$, then the mass in $r_{min}$ is
$\sim 10^5 M_{\odot}$.

In order to distinguish between the effects of reducing small-scale
power and increasing thermal velocities,
we have studied the collapse of massive
halos numerically; an 
illustration is shown in Figure (\ref{fig:pha}). 
We evolved 
small boxes (length 3 $h^{-1}$ Mpc) of $32^3$ particles, with 
particle masses of $6.86\times 10^7 h^{-1} M_\odot$. For comparative
purposes, we studied $\Lambda$CDM, and $\Lambda$WDM 
for $m_X=$ 350 eV, first with
both the appropriate linear power spectrum and the appropriate thermal velocity
distribution, and then with only the appropriate power spectrum but no
thermal velocities. We do include the particle velocities from the 
Zeldovich approximation in the initial conditions, 
with the simulations starting at $Z=40$. 
Identical phases were utilized for all three simulations.
In the Figure, we have identified a particular object for study, which
has a mass of $2\times 10^{11} h^{-1} M_\odot$ (over 3000 particles). 
We show a phase space (r,v) plot of all the particles
within a region centered on the highest density point.
The results are
shown at redshifts 8, 1 and 0. In $\Lambda$CDM, the halo shows little
evolution, having already begun to collapse at $Z=8$
and relaxed to virial equilibrium well before redshift 1.
The WDM halos both show substantial
evolution  between $Z=8$ and 1, but not much 
between 1 and 0. They are very similar to each other
at all redshifts, illustrating the unimportance of the initial thermal
velocities on these mass scales. 
The value of $\rho/<v^3>$ near the center of this
halo falls by more than four orders of
magnitude over the course of the simulation.

Let us now briefly recall the bound given by \citet{GT}
on the possible phase space density of collapsed objects. This is most
relevant to the objects with highest density and lowest velocity dispersion,
namely the dwarf spheroidal galaxies. Liouville's theorem states that 
the fine grained phase space density at the location of a fluid particle is 
constant in the absence of collisions. 
In dense regions,
the measured phase space density can decrease due
to coarse graining, but it cannot increase. 
For the canonical neutrino-like particle we are considering, the
phase space density in the initial conditions has a maximum value of
$h_P^{-3}$, where $h_P$ is Planck's constant. This is the upper bound for the
collapsed object. If one assumes that after virialization the
velocity distribution is Maxwellian, with one dimensional
dispersion
$\sigma^2$, one 
obtains a final phase space density of
 $\rho/m_X \times(2\pi \sigma^2 m_X^2)^{-{3\over2}}$. Thus the limit is
\begin{equation}
\rho < 2 m_X^4 \left( {\sqrt{2 \pi} \sigma \over h_P} \right)^3.
\label{eq:GT}
\end{equation}
In observational units this translates to 
\begin{equation}
\rho < 1.6\times 10^{-2} \, h^2 M_{\odot} {\rm pc}^{-3} 
\left({\sigma \over 
{\rm km\, s}^{-1}}\right)^3 \left({m_X \over {\rm keV}}\right)^4.
\end{equation}
With the further assumption that the core is modeled as a King profile,
with core radius $r_c$, one finds for the central density 
$\rho= 9 \sigma^2/(4 \pi G r_c^2)$ 
(see e.g. \citet{peeb},
p. 435), and thus
\begin{equation}
r_c > 32 \, \left({10 \,{\rm km\, s}^{-1} \over \sigma}\right)^{1\over 2} 
\left({{\rm keV}\over m_X}\right)^2 {\rm pc}.
\end{equation}
Draco and Ursa Minor have core radii 
of $\sim 200$  pc and $\sigma$ of 9 km s$^{-1}$ \citep{svdbbook},
so one sees that $m_X$  must be greater than about 400 eV for
warm particles to dominate these galaxies. 
But this bound is weaker than others we have described above.

\section{The Simulations}

Three simulations were carried out: $\Lambda$CDM and 
two $\Lambda$WDM models, one with warm particle mass $m_X$=350 eV 
and one with 175 eV   
($v_0=$ 0.12 and 0.048 km \,s$^{-1}$, respectively).
The cosmological
parameters are $\Omega_X=0.3$, $\Omega_\Lambda=0.7$, and $h=0.67$;
all models were normalized by $\sigma_8=0.9$ to be
consistent with observations on large scales.
Each simulation is a periodic cube of length
20 $h^{-1}$ Mpc containing $256^3$ particles,
making the particle mass $4 \times 10^7 h^{-1} M_\odot$;
a Plummer softening length of $\epsilon=1.22 h^{-1}$kpc was used.
The time step was set to be $\sqrt{0.05\epsilon/a_m}$,
where $a_m$ is the magnitude of the largest acceleration
experienced by one of the particles.
All runs were carried out with the P$^3$M code of 
\citet{Ferrell}, in the parallel version of \citet{Frederic}.

To generate the initial conditions
particles were displaced from a regular grid
in the standard Z'eldovich approximation.
The phases used for the displacements and velocities 
were identical in the three simulations.
The matter power spectrum for the LCDM model was computed using the 
COSMICS package  \citep{Bert95,MaBert95,BodBert95}.
For WDM this was modified by the transfer function $|T^X_k|^2$ 
of equation (\ref{eq:txk}) in the Appendix, suppressing small scale power. 
Warm particles were also given additional random velocities in
accordance with the appropriate Fermi-Dirac distribution.

The number and distribution of virialized halos are central to
our study. To locate collapsed objects the HOP code of 
\citet{EisHut98} was used, smoothing over 32 particles when calculating the
density at each point and using an outer density contour of 100 times
the mean.  Each halo thus identified was then considered as an
isolated object, and those particles not gravitationally bound were
removed.  Then, using the most bound particle as the halo center,
the radius $r_{200}$ of the sphere enclosing a mean overdensity of 200
was found, as well as the mass $M_{200}$ inside this radius.
Objects with $M_{200}<10^9 h^{-1} M_\odot$, i.e. with fewer than
25 bound particles, are removed from further consideration. 

These simulations do not have the resolution required to go to
the smallest observational scales, since we only resolve 
objects $\gtrsim h^{-1} 10^9 M_\odot$.
A second limitation is that the runs 
become computationally costly at low redshifts. 
Hence we shall show results for these large
boxes only
down to redshift unity. We have performed smaller ($128^3$ particles
with 3$h^{-1}$ box size) simulations
to redshift zero, for a more realistic warm particle mass 
of $1.5$  keV, which we also discuss below. 

\section{Evolution and Clustering Pattern}

Figure (\ref{fig:grid}) shows the projected density in three 
runs of $\Lambda$CDM and 
$\Lambda$WDM with warm particle masses of $175$ and $350$ eV. 
From top to bottom, the boxes are shown at redshifts $Z$ of 3, 2 and 1. 
The most obvious distinction is that the warm boxes have less
structure at $Z=3$, especially on small scales. This illustrates the
constraint posed by the requirement of early galaxy and quasar formation
in a WDM cosmology, and of course the requirement that the 
universe be re-ionized at least by $Z\sim 6$. 

The second feature of the figure is that the clustering pattern
on large scales is identical
in the three cases, as is to be expected given the
identical phases used to generate the initial conditions. 
However on
smaller scales interesting details are revealed. In the $\Lambda$CDM
model the regions between the `cosmic web' structure are filled with
small halos, which do not occur at all in the warm models. 
Small halos are formed in the warm models, but only within the 
caustic surfaces (`ribbons') of the `cosmic web'. The different clustering 
patterns are shown in more detail in Figure (\ref{fig:morph}), 
where the spatial distribution of halos of three mass
ranges are plotted. In the CDM model the small halos 
$M< 9\times 10^9 h^{-1}M_{\odot}$
fill the voids, whereas in the warm models they closely 
trace a `cosmic web' pattern of remarkably thin (and 
cold) caustic sheets.

An illustration of the typical environment of a halo 
is shown in Figure (\ref{fig:locden}). Here the local density around
each halo is calculated by measuring the total mass
within a 1 $h^{-1}$ Mpc sphere surrounding the halo center. 
The cumulative fraction of
halos as a function of the local density is plotted. 
The plot shows that half of the less massive halos 
($10^{9}$ to $9\times 10^{9} h^{-1} M_{\odot}$)
in the cold model are located in low density 
regions at less than twice the mean density. 
In the warm models, almost all such halos are in regions
denser than this.  The median local overdensity increases from
1.5 for CDM to 4 and 7 for 350 and 175 eV WDM respectively.
Results for larger halos only ($> 10^{12} M_{\odot}$) are also
plotted, revealing little difference in their mean environments 
between the three models.  
This is due in part to the fact that the halo itself 
is included when calculating the local density; centering a
sphere on a massive halo ensures that the enclosed mass is 
a few times overdense.
In the 175 eV  model, the distribution
for more massive halos is quite similar to that for small halos, showing
that smaller halos form in the same filamentary environment as
larger ones.

A blow-up of the projected density of
one octant of each cube is shown in Figure
(\ref{fig:gridoct}). 
The suppression of small scale structure in the warm models is
evident. 
The closer view of massive halos
reveals many satellite halos in the $\Lambda$CDM  model,
but very few in the warm models.
This is quantified  statistically in Figure (\ref{fig:sat}), 
which shows
the average number of satellite halos within each halo. This number defined
by measuring the distances separating each pair of halos: if this distance
is smaller than the sum of their radii (defined as the radius enclosing
mean overdensity 200), then the less massive halo is counted as a satellite
of the more massive halo. There is
strong  suppression of the number of satellites of the massive halos
in WDM, by roughly a factor of 5.
Somewhat surprising is that the 
effect is even larger for 
the 350eV warm particles than for the 175 eV particles. 
The number depends on both the rate at which satellites are accreted 
and the rate at which they are tidally destroyed \citep{CoAvVa00}.
In WDM small halos are found in denser regions;
on the other hand,  having formed later these are less dense
and thus more easily disrupted when encountering a massive halo.

Careful examination of the web structures in the central column
reveals at late times a fairly regular pattern of instability
with wavelength of maximum growth being comparable to the 
transverse ribbon dimension.  We believe this fragmentation
pattern is a real reflection of the top down instability;
this is discussed in more detail in the next section.

To summarize the findings of this section, we see 
that the bulk of the low mass halos are formed in 
the classic `top down' structure formation scenario,
via the fragmentation of large 
ribbons and pancakes.
However we see this 
taking place on relatively small scales,
within a scenario in which hierarchical `bottom-up' structure 
formation
is simultaneously occurring on larger scales.
In the next section we quantify the abundances and properties of
halos in the three simulations.

\section{Halo Abundances and Structure}

Figure (\ref{fig:pscomp}) shows the halo mass function found
in the three models.  There are two features distinguishing
the $\Lambda$WDM models from $\Lambda$CDM.
First, there is a strong suppression of low mass halos, 
consistent with the cutoff seen in the linear power spectrum.
This suppression is seen for masses below the scale set by
equation (\ref{eq:rsup}),
${4\over 3}\pi\rho_b R_S^3$, or $4\times 10^{11}$ and
$4\times 10^{12} h^{-1}M_{\odot}$ for $m_X=$350 and 175 eV respectively.
Secondly, the mass function becomes much steeper again at the lowest
masses.
However,
these low mass halos are not formed early, as the corresponding
cold dark matter halos are. Rather they are formed via pancake 
fragmentation as explained above, and therefore at lower redshifts and
with correspondingly low densities.

\citet{jfwccey00} provide a general formula which
predicts the abundances of collapsed dark matter halos
for a range of cosmologies
(the approach of \citet{smot99}, improving on the standard
Press-Schecter excursion set approach, 
gives quite similar results). 
We compare this prediction with the numerical results
in Figure (\ref{fig:pscomp}). 
$\Lambda$CDM is fit very well across the entire mass range
of the simulation. 
For the WDM models, which have a strong maximum in the power spectrum 
within the wavenumber range involved in nonlinear clustering,
the  excursion set approach fits poorly on scales below that
set by the smoothing length $R_S$.
The rms mass fluctuation $\sigma$ on scales relevant to
small masses is roughly constant, as is shown in Figure (\ref{fig:mass}).
Thus the effective spectral slope is highly negative, and well
outside the parameter space in which the analytic formulas
have been tested (see Figure (9) of \citet{jfwccey00}).
These formulae also do not predict an upturn at smaller scales,
which is hardly surprising,
since the formation of the small halos occurs via caustic fragmentation,
and has very little to do with the initial power spectrum
smoothed on the appropriate comoving mass scale.

This last point concerning caustic fragmentation is worth
examining in more detail, as it may be questioned how realistically
this has been simulated--- there may be numerical relaxation within
caustics which leads to spurious collapse.  If such relaxation is
important, it could be reduced by increasing the softening length.
As a test, the $m_X$=175 eV model was again evolved with a
Particle-Mesh only code; this new simulation is 
thus essentially collisionless.  
A mesh of 512 along a dimension was used, making the
force resolution roughly the same as the initial interparticle spacing,
and roughly a factor of 30 lower than the the resolution of 
our primary runs.
The resulting mass function is displayed in Figure (\ref{fig:pscomp})
as a dashed line; it is quite similar to the full resolution case.
Although there are somewhat fewer low mass halos (as would be
expected), the abrupt change in slope of the mass function at
$10^{11} h^{-1}M_{\odot}$ is reproduced. As objects near this
mass contain over 2000 particles, it should not be too surprising
that the reduction in resolution has little effect on such scales.
The similarity seen in the mass functions (and in visual examinations of
the projected density) of the two different resolution runs makes
us confident that the fragmentation seen in filaments is not a
numerical artifact.

A second line of approach to ensure that
the small halos are the result of caustic fragmentation is to examine
the evolution of the number density over time.
Figure (\ref{fig:zmf}) shows the cumulative mass function at 
redshifts 4 and 1 for the $\Lambda$CDM and $m_X$=350 eV models.  
For standard $\Lambda$CDM there is relatively
little change in the low mass end between these two redshifts;
most of the small halos are already in place, so the change is
less than a factor of two. On the other hand, in the WDM model 
there is an increase of roughly an order of magnitude in the
number density of lower mass objects.  The 
late formation epoch of these small halos also
supports the concept of their resulting from a top down instability.

The differences in formation history are reflected in the internal
properties of halos.
The density profile of each halo was fit to a NFW
profile \citep{NFW97}
\begin{equation}
\rho(r)= { 4 \rho(r_s) r_s^3 \over r (r+r_s)^2},
\label{eq:NFW}
\end{equation}
which is a convenient parameterization.
The core size $r_s$ is roughly the radius where the
rotation curve of such a halo should flatten, and $\rho(r_s)$ is the
value of the density at that radius,
which we term the `characteristic' density. 
This was done by computing the density in radial bins of width
3 $h^{-1}$kpc and finding the best-fit parameters $c=r_{200}/r_s$
and $r_s$ by minimizing $\chi^2$; if the reduced $\chi^2$ gave
less than a 90\% probability for the fit then that halo was
not used in the following analysis.

Figure (\ref{fig:halofits})
shows the resulting concentration parameter $c$, $r_s$, and
$\rho_s$ as a function of halo mass for each of the models studied. 
For the lowest masses the differences are quite substantial.  
The WDM halos
show lower densities and central concentrations
than the corresponding cold dark matter ones.  
Thus the WDM halos are in better agreement with observations
of nearby dwarf galaxies than standard CDM \citep{vdBRDB}.
While smaller, these differences
persist up to higher mass halos formed in the normal
(bottom up) fashion.  In the normal CDM model the central parts
of high-mass halos are from accreted low-mass, high-density
halos \citep{SCO00}; in WDM such accreted objects are rare.
\citet{CoAvVa00}, using a
warm particle mass of 604 or 1017 eV, found that
lower-mass satellites found around Milky-Way sized halos
have concentration factors lower by a factor of 2 compared to 
$\Lambda$CDM.  This appears to be in good agreement with
Figure (\ref{fig:halofits}).

\section{A Higher Warm Particle Mass $m_X$}

The WDM simulations discussed above are for particle masses
too low to reconcile with the limits discussed in Section I,
which indicate $m_X\gtrsim 1$ keV.  While they are
useful for identifying qualitative trends, one may ask if
the differences with standard CDM discussed above persist to
higher particle masses.  To test this, an additional pair
of simulations were made;  one was $\Lambda$CDM with the
same cosmological parameters as before, and the other was
$\Lambda$WDM with $m_X=1.5$ keV. The box size was chosen
to be 3 $h^{-1}$Mpc, with $N=128^3$ particles.  These runs
were carried out and analyzed in the same manner as described in Section 5;
because of the smaller number of particles they could be 
evolved to redshift zero. For such a small box, it is
inevitable that there are sizable finite size effects, and these
are visible 
in the orientation of the largest pancake.

The trends seen for lower warm particle masses are also present
in this 1.5 keV run.  Figure (\ref{fig:1p5kmorph}) shows the
positions of halos identified in the two simulations;
shown are halos with bound masses larger than 
$2.7\times 10^7 h^{-1}M_{\odot}$ (25 particles).
The number of low mass halos is suppressed in the WDM model
by a factor of over three, and it can be seen that small
halos trace the caustic structures connecting the largest
ones.  This is in contrast with small CDM halos, which occur in
voids as well.  
The WDM halos tend to be found in denser environments as well. 
Using a sphere of $300h^{-1}$kpc to define the local overdensity,
only 20\% of the WDM halos are in regions with local overdensity
less than unity, while 40\% of the CDM halos are.
The larger WDM halos have fewer satellites:
only 17 satellites are found around 
the ten largest WDM halos (with masses $\gtrsim 10^{10}
h^{-1}M_{\odot}$), while the ten largest CDM halos have 97. 
The WDM halo profiles are also quite distinct from CDM. 
Fitting with the NFW profile, the mean concentration parameter
is lower by a factor of 2.5, and the mean core radius is larger
by a factor of 3. The overdensity at $r_s$ 
has mean $\sim 2\times 10^5$
in $\Lambda$CDM with a few halos above $10^6$; 
in WDM the mean is  $\sim 2\times 10^4$ with $\sim 10^5$ as the maximum. 

There is definitely room for more detailed study, but this pair
of runs indicates that the trends described in our larger 
runs do persist as the warm particle mass is increased.

\section{Conclusions}

We have identified several key observational features of WDM
clustering which serve to distinguish it from cold dark matter.
The principal difference is due to the `top down' formation of structure 
below a characteristic mass scale, which we can roughly quantify as
\begin{equation}
M_{S} =  1.0\times 10^{10} \left({\Omega_X\over.3}\right)^{1.45} \left({h\over 
.65}\right)^{3.9}
\left({{\rm keV}\over m_X}\right)^{3.45} h^{-1} M_{\odot}.
\label{eq:masssc}
\end{equation}
Below this mass scale, objects are formed primarily by the fragmentation 
of pancakes and ribbons. They are rarer and 
considerably less dense than halos of
the same mass in CDM. And their spatial distribution is very different -
they are concentrated in sheets and ribbons running between
the massive halos, an effect which has been noted for some time for dwarf 
galaxies in the local universe \citep{pjep00,peeb01}.
Finally, the dwarf galaxies form later than bright galaxies, 
contrary to the hierarchical picture of clustering in CDM theory.
Again WDM seems to be in better agreement with the observations on
this point \citep{metc}.

This work points to the need for the following observational tests
to verify or refute the predicted signatures, namely:

$\bullet$ Observations revealing the ionization history
and Lyman $\alpha$ cloud distribution at redshifts higher than 6. 

$\bullet$ Observations revealing the history of galaxy formation 
and checking whether low mass 
galaxies formed later than
higher mass galaxies, as in WDM, or earlier as in CDM.

$\bullet$ Determination of the 
abundance, masses, density profiles   and clustering pattern of dwarf galaxies 
relative to large galaxies.
A recent study \citep{zab} indicates that the dwarf to giant ratio
increases with increasing density, consistent with our findings here.
Likewise the apparent absence of dwarf systems in the voids
noted by \citet{pjep00,peeb01} needs to be 
be carefully
quantified and compared with CDM and WDM simulations. 

$\bullet$ High accuracy observations measuring the inner parts of
galaxy rotation curves and comparing them with detailed N body simulations
of $\Lambda$WDM and $\Lambda$CDM. 

$\bullet$
Observations of the abundances of satellites in 
galaxies other than the Milky way. 
Especially important is a determination of the characteristic densities
of dwarf dark matter dominated systems, as compared to dwarf halos
in simulations.

It is clear that a great deal also needs to be done in simulating WDM,
to better quantify the qualitative features we have noted here. 
This will require higher resolution codes and considerable amounts
of computation time.


\acknowledgments

We would like to thank R. Battye, G. Efstathiou, R. Lopez, J. Peebles,
M. Rees, and P. Steinhardt for useful discussions.
We also thank S. Colombi for a detailed and useful referee's
report, and R. Barkana for many useful suggestions.
Special thanks are owed to C-P. Ma who kindly provided
Boltzmann code results.

This research was supported by NSF Grant
AST-9803137 (under Subgrant 99-184), and the NCSA Grand Challenge
Computation Cosmology Partnership under NSF
Cooperative Agreement ACI-9619019, PACI Subaward 766.
The work of NT
was supported by a PPARC (UK) rolling grant, and PPARC and HEFCE (UK)
grants to
the COSMOS facility. NT thanks the
ITP (Santa Barbara) for hospitality while this paper was being
completed.



\appendix

\section{Appendix: Linear Theory}

In this appendix we review the standard results concerning thermal 
relic warm dark matter and then discuss a simplified approach to 
understanding damping by free streaming, which illustrates the 
physics and tells us how the power spectrum should scale as
the mass of the warm particle, and the cosmological parameters
are varied.  We explain the calculation of the smoothing length
$R_S$ given in equation (\ref{eq:rsup}) above.

For a thermal relic $X$ which decouples when relativistic,
the abundance relative to photons is
\begin{equation}
{n_X\over n_{\gamma}} = \left({43/ 4\over  g^*_{dec}}\right)\left({2\over 
11/2}\right)  {g_X\over 2}
\label{eq:one}
\end{equation}
where $g^*_{dec}$ is the effective number of relativistic
species present at $X$ decoupling, contributing the entropy
of the universe. Bosonic degrees of freedom contribute unity to 
$g^*$, fermionic degrees of freedom contribute ${7\over 8}$. 
The first factor in (\ref{eq:one}) represents the dilution
of $X$ particles relative to photons as the $g_{dec}$ degrees
of freedom present after $X$ decoupling decay into photons,
electrons, positrons and neutrinos (which together have 
$g^*={43\over 4}$). The second factor is the further dilution 
of $n_X/n_\gamma$ 
when electrons and positrons annihilate. Finally, the last
factor is the number of degrees of freedom $g_X$ contributing
to the number density. Bosons contribute $g=1$ but
fermions $g={3\over 4}$. This is divided by the number of
degrees of freedom for photons, which is 2. A light neutrino
species contributes $1.5$ to $g$, and it is conventional to use this
as the fiducial value for $g_X$ for the
warm particle.

The abundance of $X$'s fixes their density through $\rho_X=m_X n_X$.
This yields 
\begin{equation}
\Omega_X h^2 \approx  {115\over g_{dec}}\, {g_X\over 1.5}\, {m_X\over {\rm 
keV}}.
\end{equation}
As mentioned in Section 2, if entropy producing processes occur subsequent to 
$X$ decoupling, they act in a manner as to mimic an increase in 
$g_{dec}$. 

The second quantity of interest is the velocity dispersion of
the $X$ particles, since this is what causes the smearing of
small scale primordial perturbations.
If the $X$ particles decouple when relativistic, the
distribution function  $(e^{(p/ T_X)} + 1)^{-1}$ remains 
constant until gravitational clustering begins. All particle momenta
scale as $a^{-1}$ which we can describe by scaling $T_X$ accordingly.
When the particles become non-relativistic we can use
$p=m_X v$, with $v$ their speed. The distribution function is
then $(e^{(v / v_0)}+ 1)^{-1}$, with 
\begin{equation}
v_0(Z) \approx .012\, (1+Z) 
\left({\Omega_X\over.3}\right)^{1\over 3} \left({h\over .65}\right)^{2\over 3}
\left( {1.5\over g_X}\right)^{1\over 3} 
\left({{\rm keV} \over m_X}\right)^{4\over 3} {\rm km\, s}^{-1},
\label{eq:veq}
\end{equation}
at redshift $Z$. The  
{\it rms} velocity is $3.571v_0$. 

A crude estimate of the free streaming lengthscale is obtained by
calculating the comoving distance a particle moves after
matter-radiation equality. This is proportional to (\ref{eq:veq})
evaluated at $Z_{eq} = 2.39\times 10^4 \Omega_M h^2$, times the
comoving horizon scale at that time, $16.0 /(\Omega_M h^2)$ Mpc
(for three light neutrinos).
Thus one obtains the scaling given in (\ref{eq:geq}), from 
\citet{bs} and \citet{bbks}. As mentioned in the introduction
however this scaling is
incorrect 
due to additional streaming in the
radiation era, as we shall now discuss. 

For the masses $m_X$ we discuss here, most 
of the smoothing of the initial perturbations occurs when
the particles are nonrelativistic. The free streaming may
then be understood via an analytic solution of the Gilbert equation
obtained by treating the background radiation as uniform,
and making a certain approximation to the Fermi-Dirac distribution
\citep{bkt}. This approximation is useful for 
revealing the main
parameter dependence of the transfer function.

The
perturbation in each plane wave mode of $X$ 
is governed by Gilbert's 
equation (for a derivation see \citet{bkt})
\begin{equation}
\delta_{X,\bf{k}}(z)= \delta^{nog}_{X,\bf{k}}(z,z_0)+ 
6\int_{z_0}^z dz' \delta_{X,\bf{k}}(z') {F(z,z')\over 
\left[1+A^2 F(z,z')^2\right]^2}
\label{eq:del}
\end{equation}
where $\delta^{nog}_{X,\bf{k}}(z,z_0)$ is a solution to
the free streaming equation in the absence of self-gravity.
$F(z,z')= {\rm ln} (1+z'^{-1})-{\rm ln}(1+z^{-1})$, 
where $z= \tau /(4 \tau_*)$, and $\tau$ is the conformal time.
The scale factor is chosen to be unity at 
matter-radiation equality, and is given by 
 $a=4(z+z^2)$
 \citep{bkt} (Note that $z$ is not the redshift). 
In (\ref{eq:del}), the quantity  $A=k\tau_*v_0(Z_{eq})$
where $k$ is the comoving wavenumber, and the
comoving value of the scale
$(1+Z_{eq}) \tau_* = 19.4 \Omega_M^{-1} h^{-2}$. The comoving horizon 
scale given earlier is $2(\sqrt{2}-1)$ times this scale.
Finally, $v_0(Z_{eq})$ is given in equation (\ref{eq:veq}).

Equation (\ref{eq:del}) is derived by approximating the 
distribution function  $(e^{p\over T}+ 1)^{-1} $ 
as $e^{-{p\over T}}$ times an appropriate normalization
factor \citep{bkt}. In this approximation bosons and fermions 
behave identically.

Equation (\ref{eq:del}) shows the scale 
$\tau_*v_0(Z_{eq})$ entering, in comoving units. This 
is the speed of the warm particles times the horizon 
at matter-radiation equality. This leads to the estimate of 
$R_{>eq}$ in (\ref{eq:geq}). However, the Gilbert equation requires initial
conditions, and specification of the non-gravitating 
solution $\delta^{nog}_{X,\bf{k}}(z,z_0)$. This is where additional
parameter dependence enters. The point is that no growth of perturbations
occurs whilst the $X$ particles are relativistic, because the
Jeans length is of order the horizon scale. When the $X$ particles
slow down, the Jeans length starts increasing,
reaches its maximum, and thereafter 
decreases. The Gilbert equation only applies after the maximum
Jeans length is attained, which is when the $X$'s become nonrelativistic.
The $X$ particles' 
rms speed falls to $\sim 0.3c $ when $z$ becomes greater than
\begin{equation}
z_0 \approx
.02 \, (\Omega h^2/m_X)^{4\over 3} (1.5/g_X)^{1\over 3}
\label{eq:zo}
\end{equation}
(with $m_X$ in keV),
which we adopt as the initial value in the Gilbert equation.
Note that this combination of parameters 
differs from that entering  $R_{>eq}$ in (\ref{eq:geq}).

The free streaming 
solution $\delta^{nog}_{X,\bf{k}}(z,z_0)$ should be calculated 
using the 
full relativistic equations, by integrating 
the perturbations 
across horizon crossing. However, no
additional cosmological parameter dependence enters this phase 
since it is deep in the radiation era.
If we are only interested
in the suppression of structure relative to that in cold dark matter,
it is reasonable to take a simple isotropic ansatz (i.e. a perturbation
of the local temperature) for 
$\delta^{nog}_{X,\bf{k}}(z,z_0)$ and then compare its evolution 
with that predicted from the same equation 
for the $k=0$ mode, which is unaffected by streaming.
In the same approximations as used above, we find
\begin{equation}
\delta^{nog}_{X,\bf{k}}(z,z_0) \propto {1-{1\over 3} A^2 F(z,z_0)^2 \over 
\left(1+A^2 F(z,z_0)^2\right)^3}.
\end{equation}
With the source fully specified, 
equation (\ref{eq:del}) is easily iterated numerically into the matter
era where the amplitude of the growing mode density perturbation 
is compared to that for $k=0$ to obtain the transfer function.

The smoothing kernel depends on the
smoothing scale $R_{>eq}$ in (\ref{eq:geq}), and $z_0$ in (\ref{eq:zo}).
Dimensional analysis then yields a smoothing scale
$R_{>eq} f(x) $ with $f(x)$  an arbitrary function and $x=
(\Omega h^2/m_X)^{4} (1.5/g_X)$. The results of the Gilbert
equation code are well fitted by
\begin{equation}
T^X_k= \left(1+(\alpha k)^2\right)^{-5}
\label{eq:txko}
\end{equation}
with $\alpha= 0.05 \, (\Omega_X/.4)^{.15} (h/.65)^{1.3} 
({\rm keV}/m_X)^{1.15} (1.5/g_X)^{.29} $ and $k$ in $h$ Mpc$^{-1}$, i.e.
$f(x) = .05 x^\beta$ and $4\beta\approx -0.183$. This additional
dependence, which weakens the decrease of smoothing length as 
$m_X$ is increased, is due to the
fact that higher mass particles 
become non-relativistic sooner and therefore go through
a longer period of free streaming in the radiation era.

A full Boltzmann code calculation \citep{Ma},
yields a fit 
\begin{equation}
T^X_k= \left(1+(\alpha k)^{2 \nu}\right)^{-5/\nu}
\label{eq:txk}
\end{equation}
with 
\begin{equation}
\alpha=0.048  \, (\Omega_X/.4)^{.15} (h/.65)^{1.3} 
({\rm keV}/m_X)^{1.15} (1.5/g_X)^{.29}, 
\label{eq:filt}
\end{equation}
and $\nu=1.2$, 
which is accurate to a few per cent over the relevant
range of $k$. The scaling, and the fitting function,
are illustrated in Figure (\ref{fig:scaling}).
The warm dark matter power spectrum to be input into 
N body simulations is then given by that for cold dark
matter
multiplied by $|T^X_k|^2$. The particle velocity distribution
is given in (\ref{eq:veq}).





\clearpage



\epsscale{0.7}
\plotone{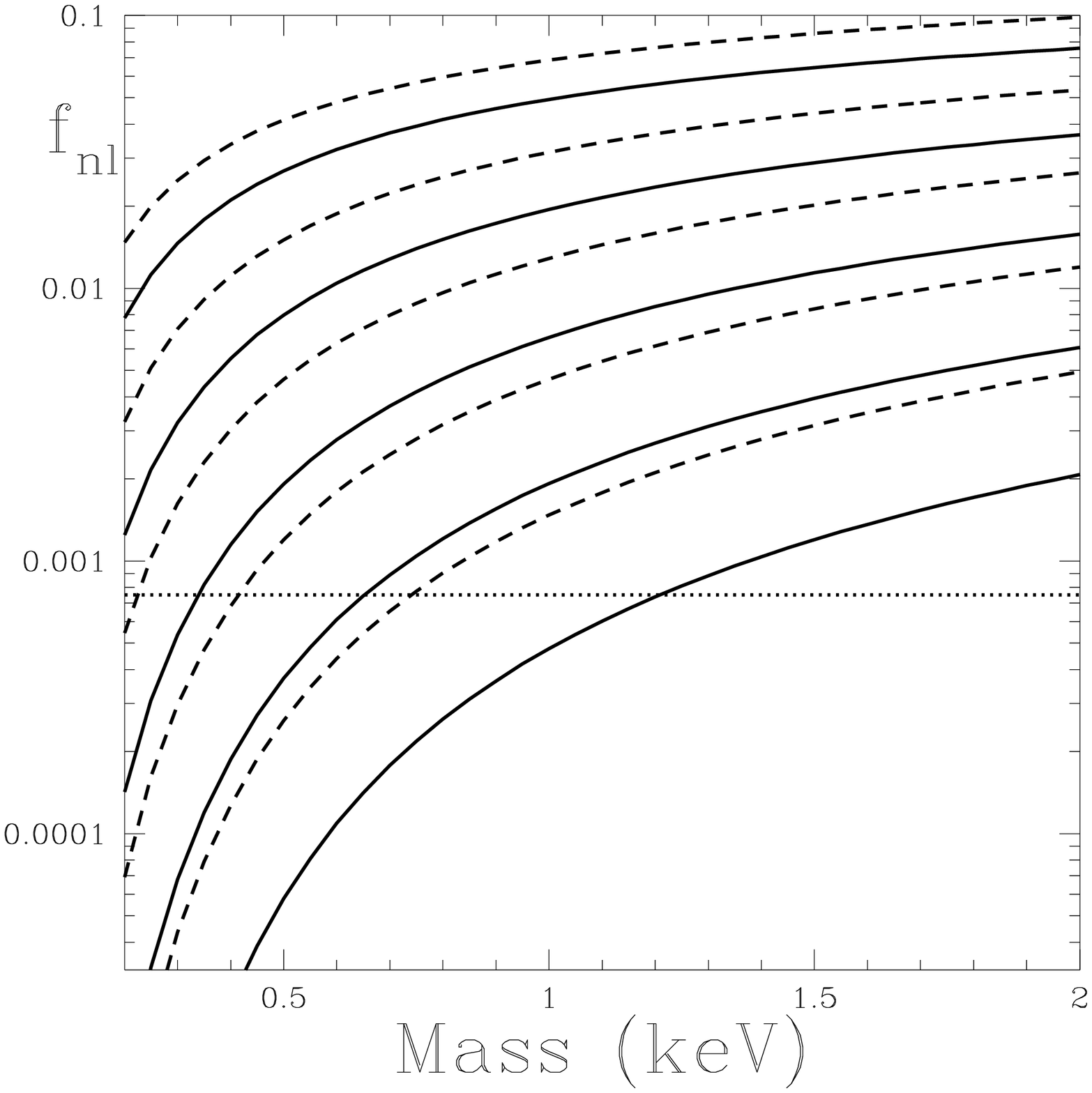}
\epsscale{1.0}
\figcaption{ The collapsed fraction $f_{nl}$, defined to be  
the fraction of matter in the universe which has 
attained a linear theory perturbation amplitude corresponding to
collapse to zero radius in the spherical collapse model.
The fraction $f_{nl}$ is 
plotted at redshifts 7, 6, 5, 4, 3 (lower to upper curves) 
for warm particle masses from 0.2 to 2 keV. The solid curves 
are for
the theory normalized to $\sigma_8=0.9$ and the dashed curves for
$\sigma_8=1.0$. The horizontal line shows an estimate
 of the fraction of mass which must be collapsed
 in order to re-ionize the universe. This is
 based on the ionizing radiation released by typical stellar
 populations today, multiplied by
 an estimated  factor of $3$
 to account for inefficient propagation and
 recombination.
The observation of a quasar at $Z=5.8$ whose spectrum exhibits little 
absorption due to neutral hydrogen thus sets a 
lower limit on the mass of the warm particle,  $m_X \gtrsim \,400$ eV
or $700$ keV for $\sigma_8=1.0$ or $0.9$ respectively.
Cosmological parameters
$\Omega_M=0.3$, $\Omega_\Lambda=0.7$ and $h=0.65$ are assumed.
\label{fig:ion} }

\epsscale{0.8}
\plotone{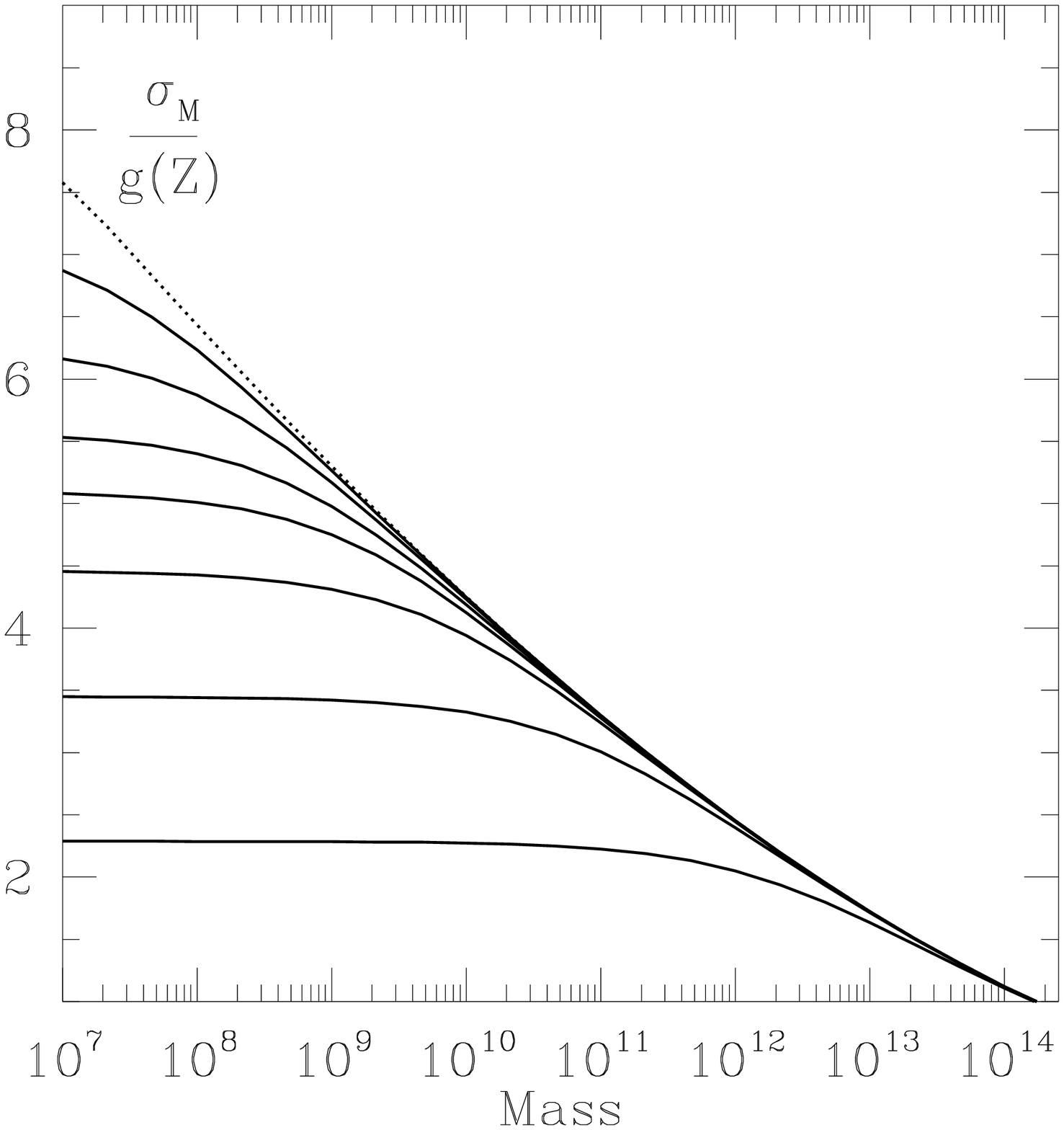}
\epsscale{1.0}
\figcaption{ The rms mass 
fluctuation $\sigma_M$ in linear theory, in a spherical
top hat window, given 
in units of the linear theory growth factor $g(Z)$ which is normalized to
unity today. The solid lines show $\sigma_M$ against the comoving mass in the
window 
for WDM particle masses of $0.2, 0.5, 1.0, 1.5, 2.0$,
$3.0$ and 5.0 keV (bottom to top). 
The theories are normalized to $\sigma_8=0.9$,
consistent with an assumed  $n=1$ primordial spectrum and the COBE results. 
The dotted line shows the same quantity $\sigma_M/g(Z)$ for CDM.
Cosmological parameters
$\Omega_M=0.3$, $\Omega_\Lambda=0.7$ and $h=0.65$ are assumed.
\label{fig:mass} }

\plotone{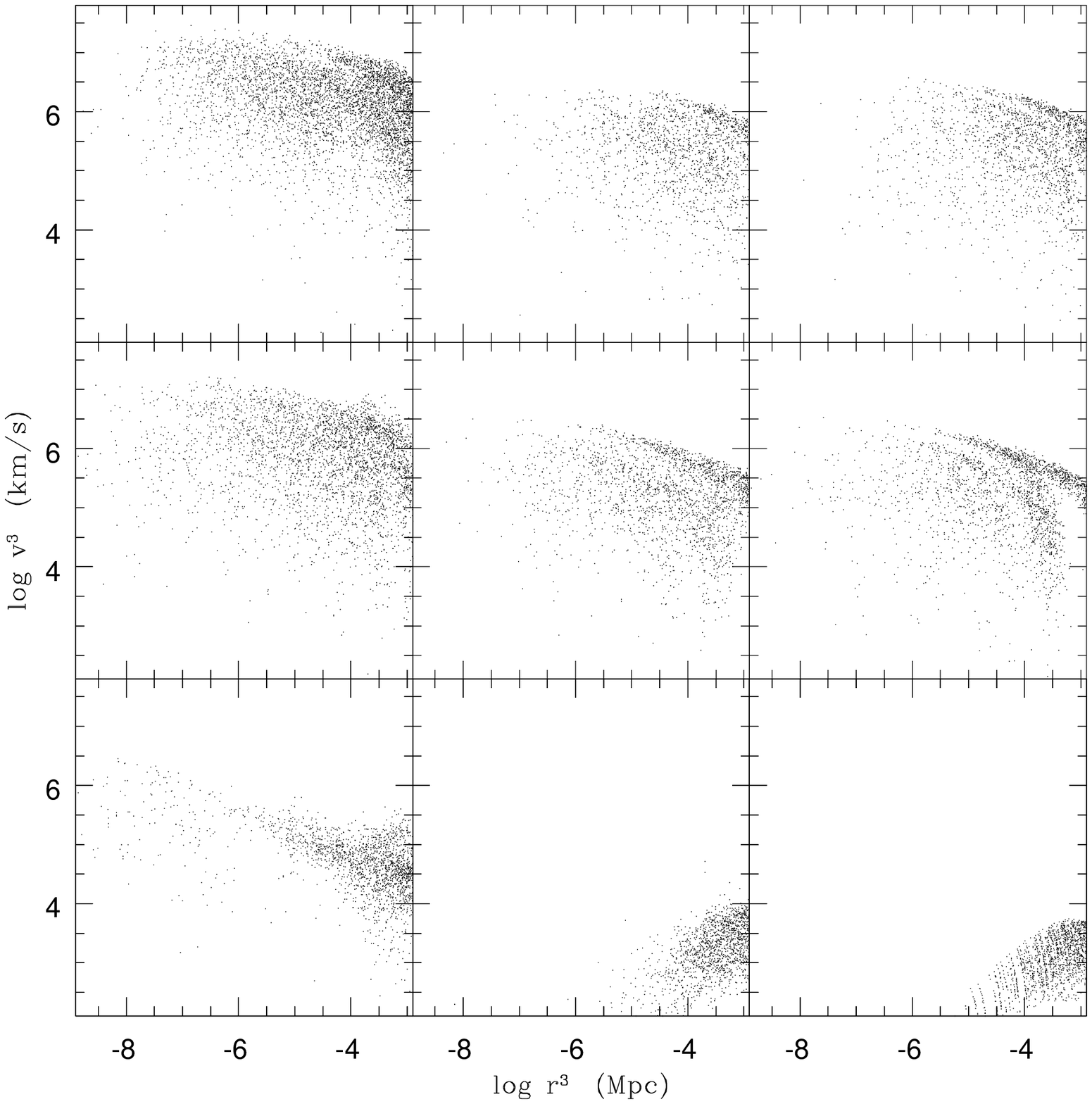}
\figcaption{The evolution of the particle distributions
in phase space. A small halo of mass $2\times 10^{11} h^{-1} M_\odot$
has been selected for comparative study in 
(left to right) $\Lambda$CDM, $\Lambda$WDM, and 
$\Lambda$WDM power spectrum but without thermal velocities.
From bottom to top: $Z=$ 8, 1, and 0.
\label{fig:pha} }

\plotone{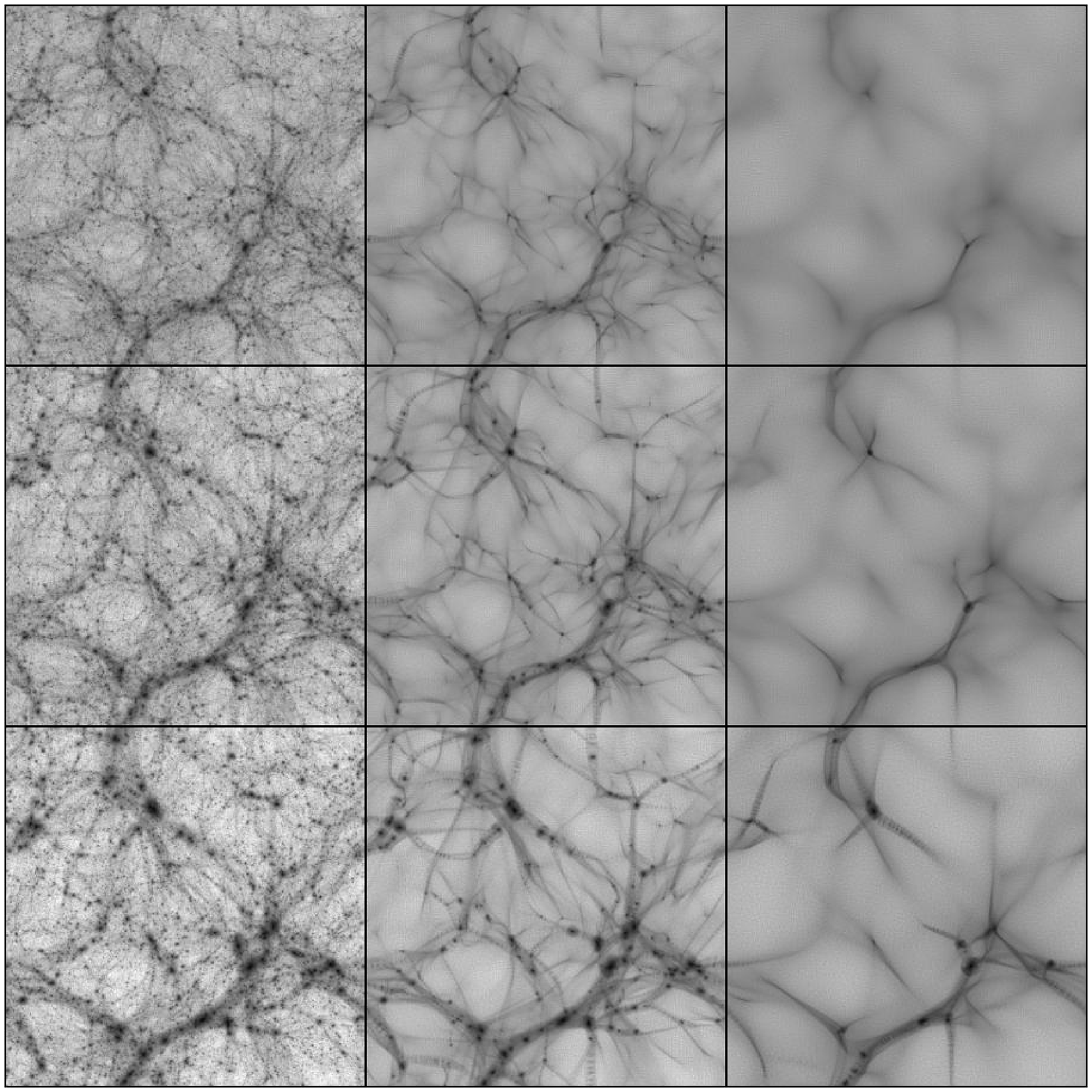}
\figcaption{
Projected density of 20 $h^{-1}$Mpc boxes, on a logarithmic scale
of surface density.
Left to right: $\Lambda$CDM, 
$m_X$=350 eV and 
$m_X$=175 eV  $\Lambda$WDM. 
Top to bottom: redshift $Z=$3, 2, and 1.
A simulation with $m_X \sim 1$ keV would have an appearance 
intermediate between the left and central columns.
(A higher resolution version of this Figure is available at the
web site referred to in the introduction.)
\label{fig:grid}  }

\plotone{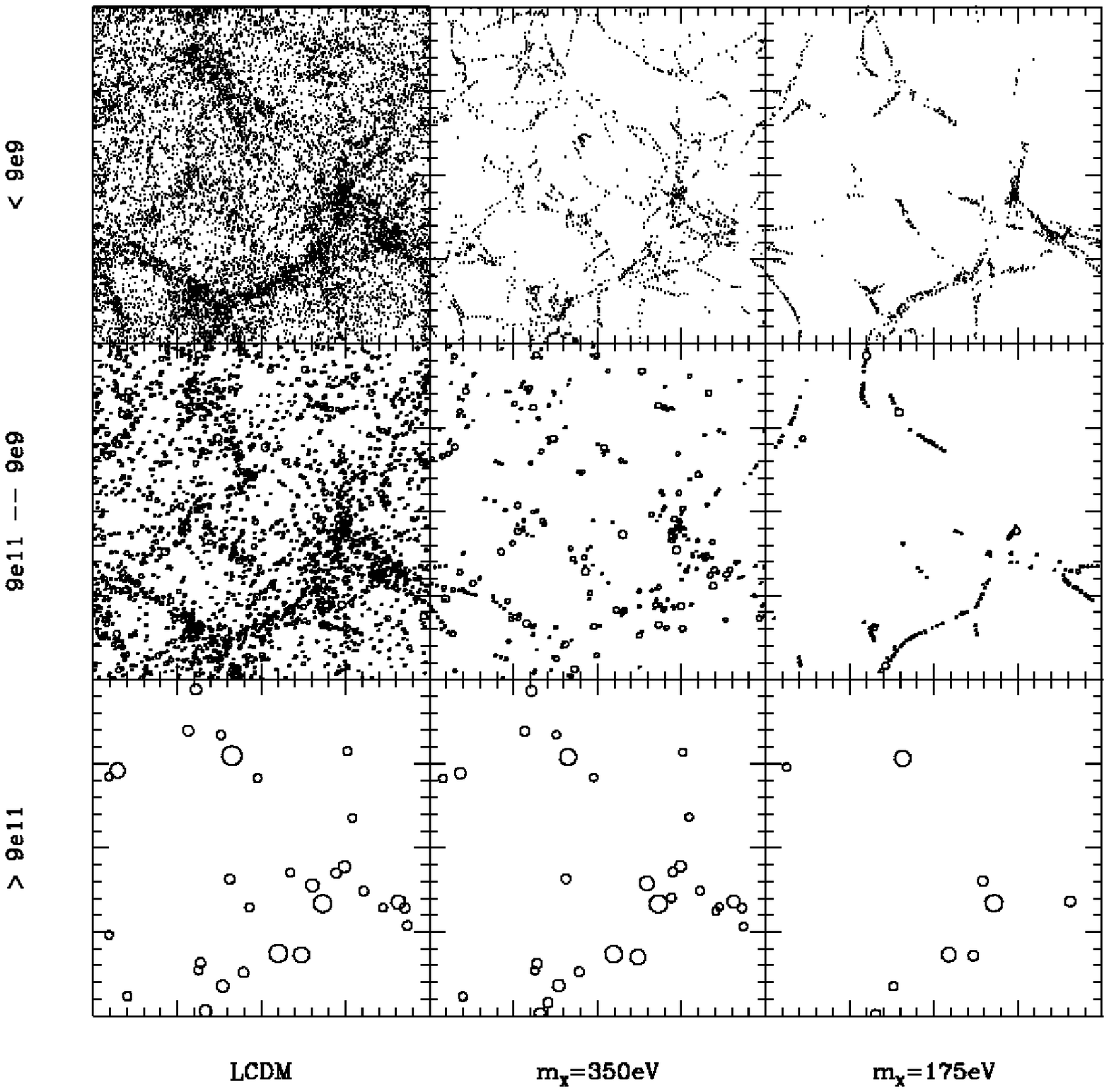}
\figcaption{Position of gravitationally bound halos at redshift $Z=1$,
corresponding to the top row of Figure (\ref{fig:grid}).
Bottom to top: high ($M_{200}>9\times10^{11} h^{-1}M_\odot$), 
intermediate 
($9\times10^9 h^{-1}M_\odot < M_{200} < 9\times10^{11}h^{-1}M_\odot$),
and low mass
($10^{9}h^{-1}M_\odot < M_{200} < 9\times10^{9}h^{-1}M_\odot$)
halos.
Left to right: $\Lambda$CDM , $m_X=350$ eV and $m_X=175$ $\Lambda$WDM. 
The radius of each circle is $r_{200}$.
In WDM, 
the formation of low mass objects is in ribbon-like structures
connecting higher mass halos. Note the almost complete absence of
small halos outside these regions, while the highest mass halos
(bottom row) are quite similar to LCDM.
\label{fig:morph} }

\epsscale{0.8}
\plotone{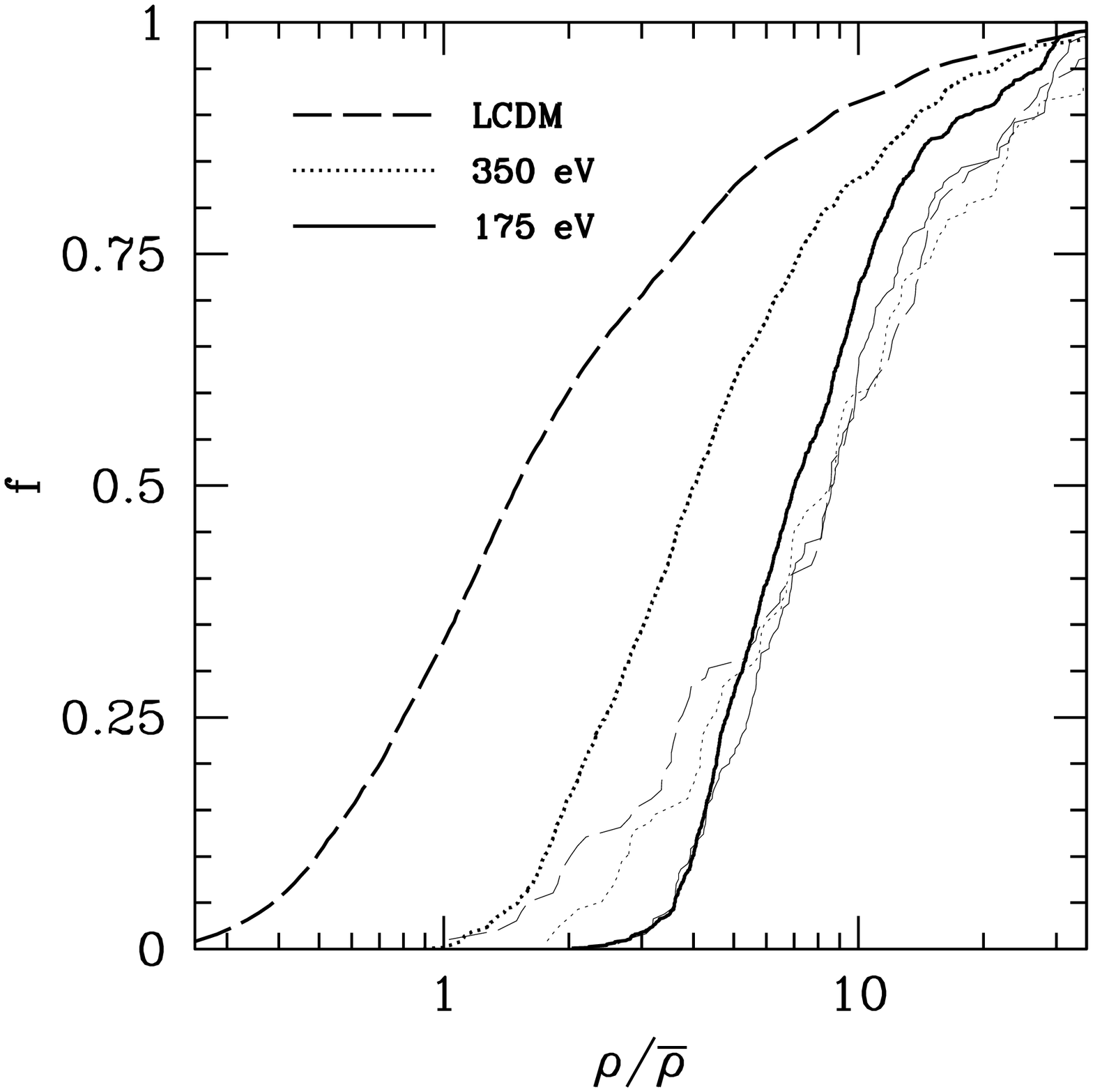}
\epsscale{1.0}
\figcaption{The mass density in the vicinity of halos. The plot 
shows the cumulative fraction of halos as a function of the mean
density in a 1 $h^{-1}$ Mpc radius sphere centered on the halo.
Thick lines are for halos with mass
($9\times 10^9>M>10^9 h^{-1}M_{\odot}$), 
thin ones for only mass greater than
$10^{12} h^{-1}M_{\odot}$. 
One sees that the WDM models have far fewer
halos in low density neighborhoods: they form instead in higher
density caustics. For high mass halos the differences in 
environment are relatively small.
\label{fig:locden} }

\plotone{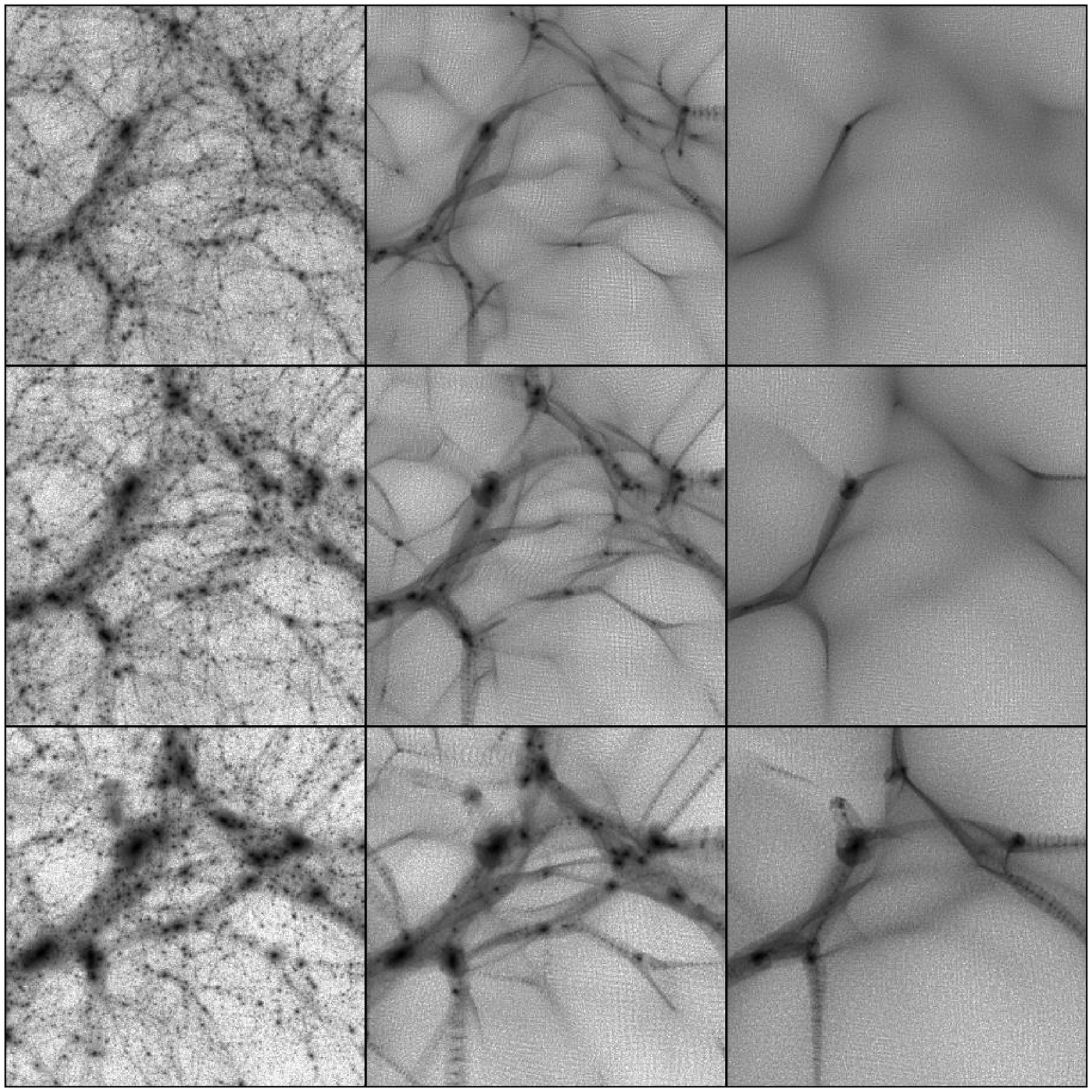}
\figcaption{Projected density in one octant of the $20 h^{-1}$ Mpc
cubes shown above. The small scale structure present in the
CDM model is clearly visible, as is the relative suppression of 
satellite galaxies about the larger halos in the warm models. 
A higher resolution version of this Figure is available at the
web site referred to in the introduction.
\label{fig:gridoct} }

\plotone{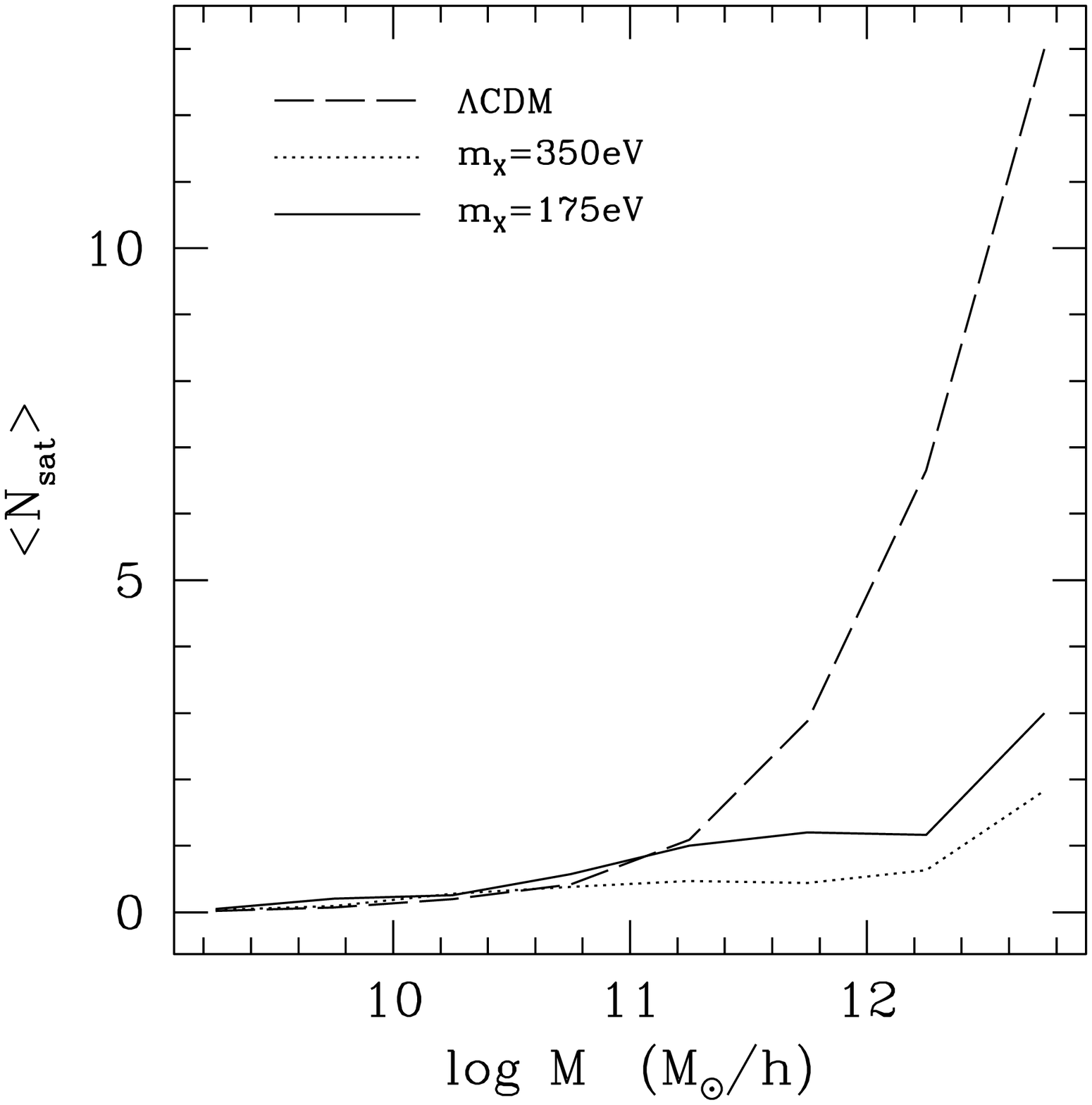}
\figcaption{Average number of satellites, as a function of the mass
of the parent halo.
See text for method of satellite identification.
\label{fig:sat} }

\plotone{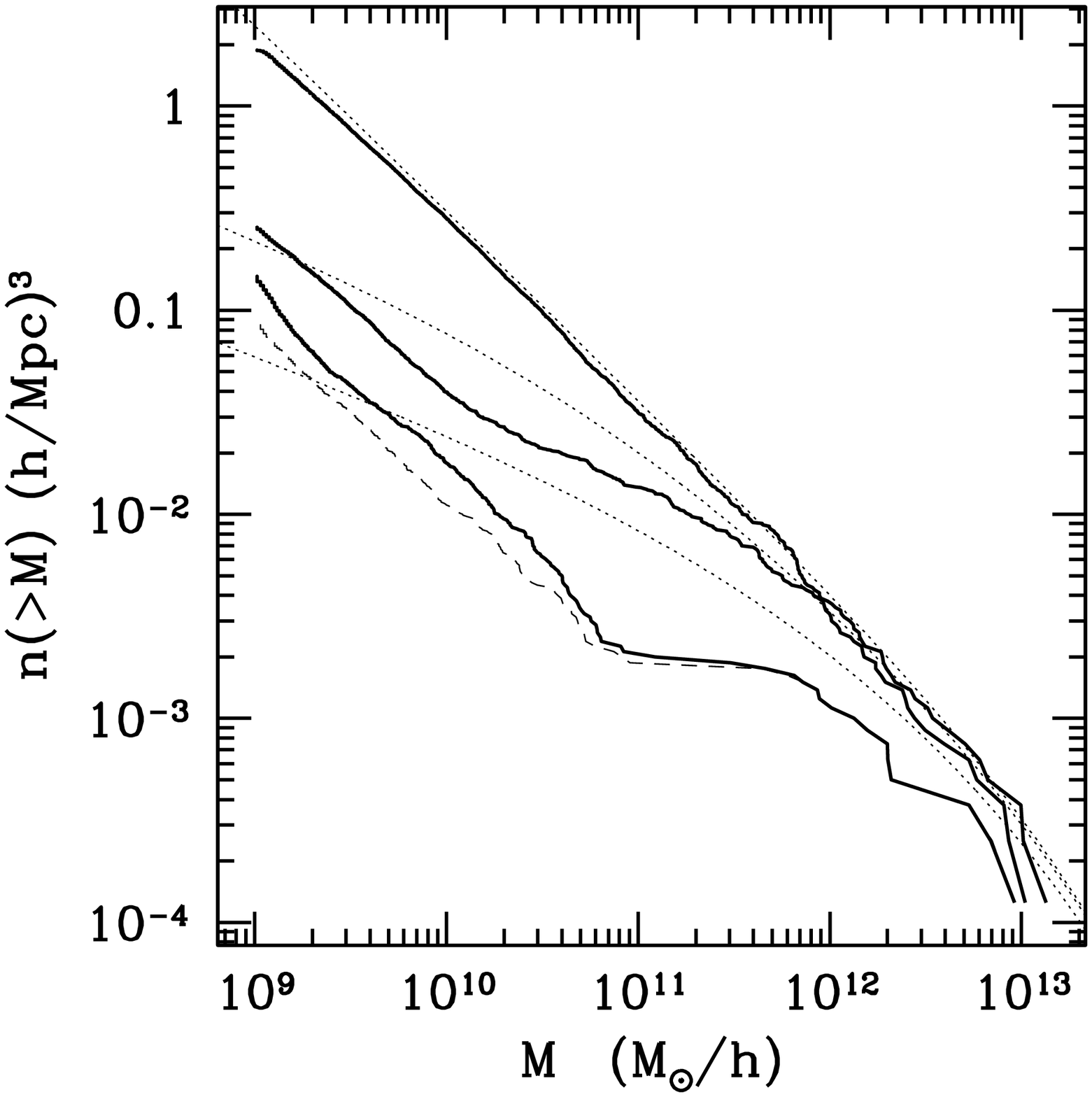}
\figcaption{Number density of halos greater than a given mass 
for the three models
(thick lines). Top to bottom:
$\Lambda$CDM, $m_X=350$eV ($v_0=0.048$ km s$^{-1}$) 
and $m_X=175$eV ($v_0=0.12$ km s$^{-1}$) $\Lambda$WDM.
The thin dashed lines are
predictions from the formula of \citet{jfwccey00}.
The dashed line is a repeat of the $m_X=175$eV run using
a PM only code.
\label{fig:pscomp} }

\plotone{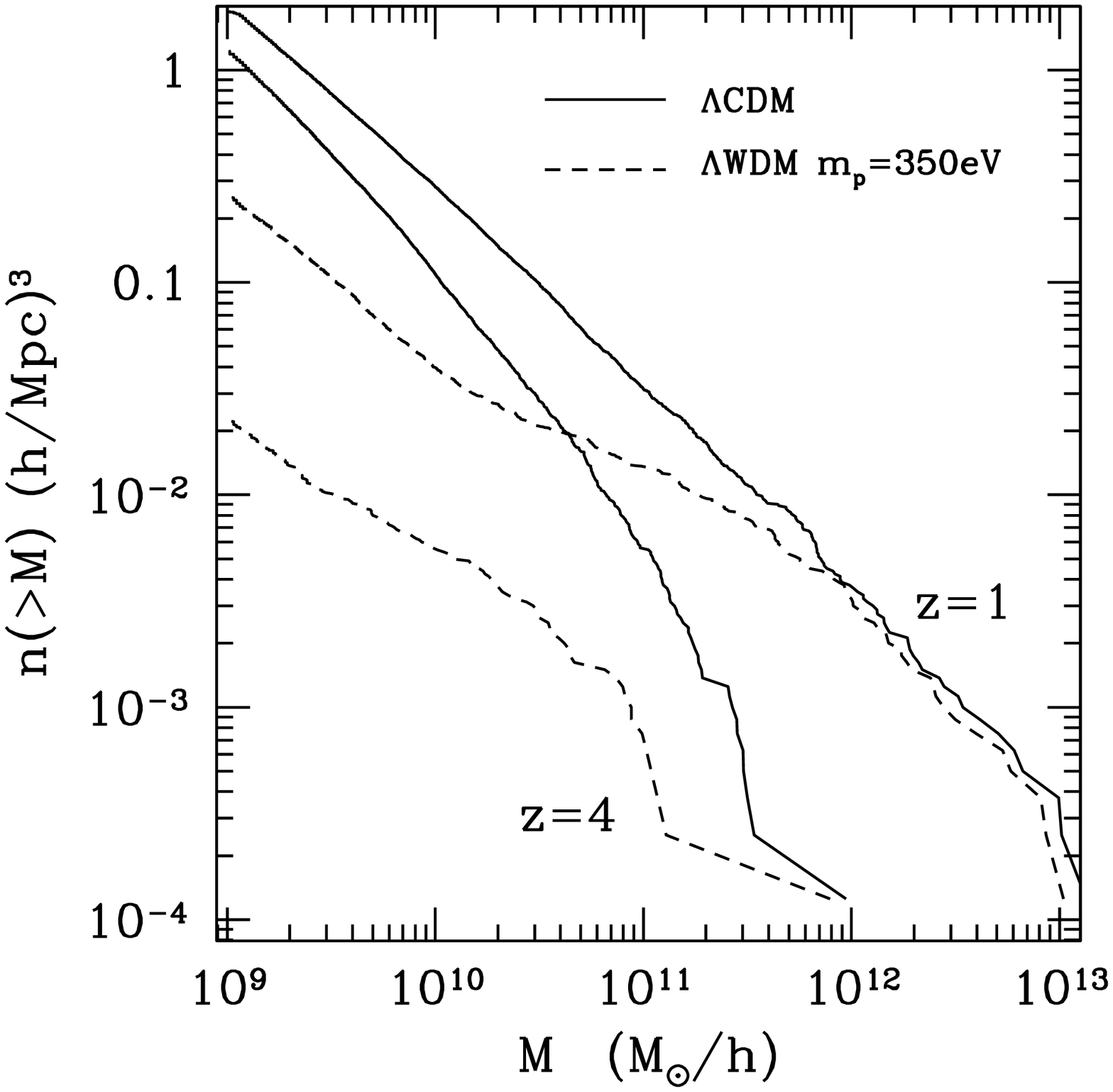}
\figcaption{Cumulative mass function at redshifts $Z=4$ and $Z=1$.
Solid lines: $\Lambda$CDM; dashed lines: $m_X=350$eV $\Lambda$WDM.
\label{fig:zmf} }

\plotone{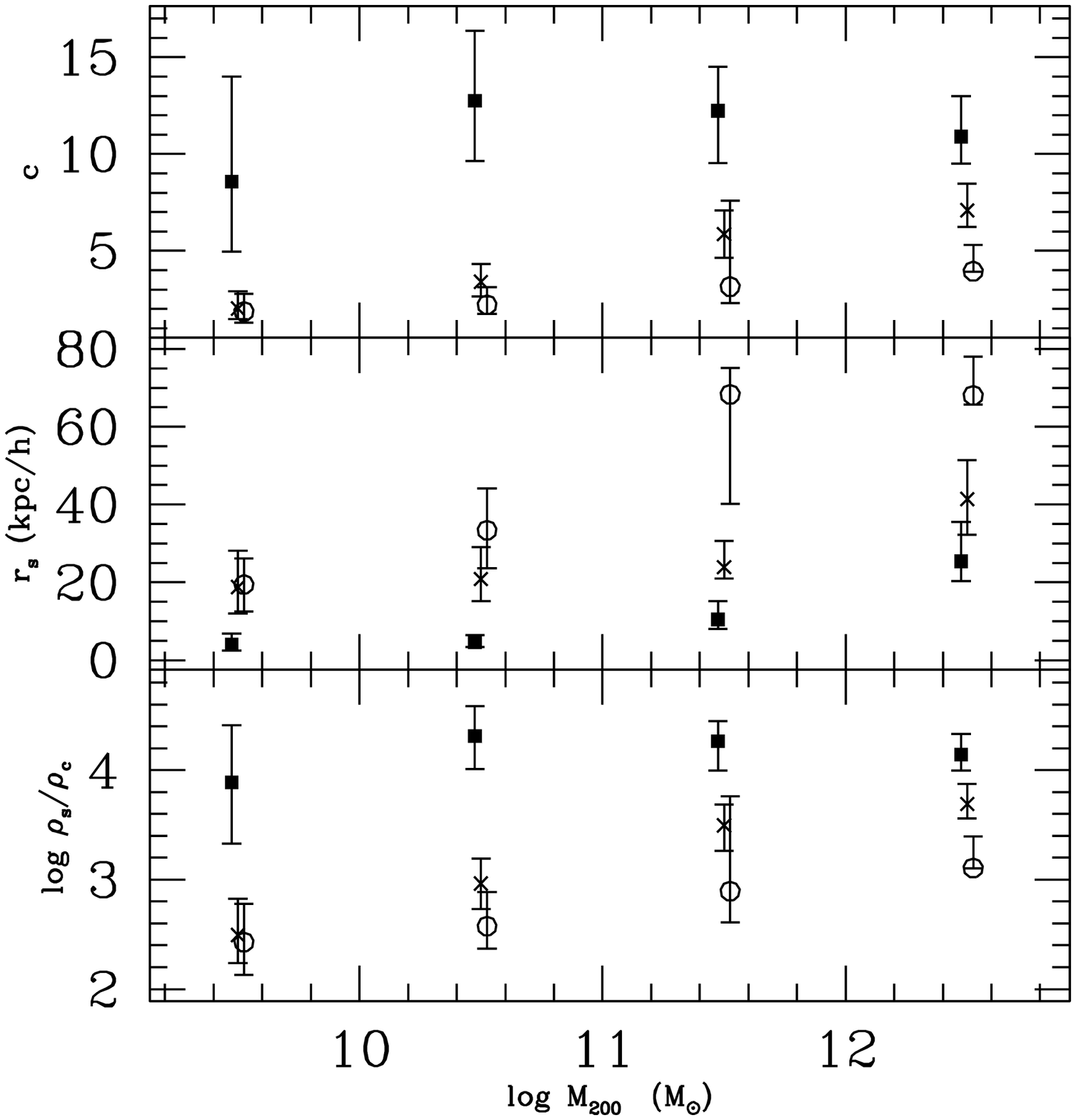}
\figcaption{Mean values for the NFW concentration parameter $c=r_{200}/r_s$, 
core radius and core density (defined in equation \ref{eq:NFW}) 
as a function of mass $M_{200}$.
{\it Filled squares:} $\Lambda$CDM. {\it Crosses:} $m_X=350$eV 
($v_0=0.048$ km s$^{-1}$) $\Lambda$WDM.
{\it Circles:} $m_X=175$eV ($v_0=0.12$ km s$^{-1}$) $\Lambda$WDM.
\label{fig:halofits} }

\plotone{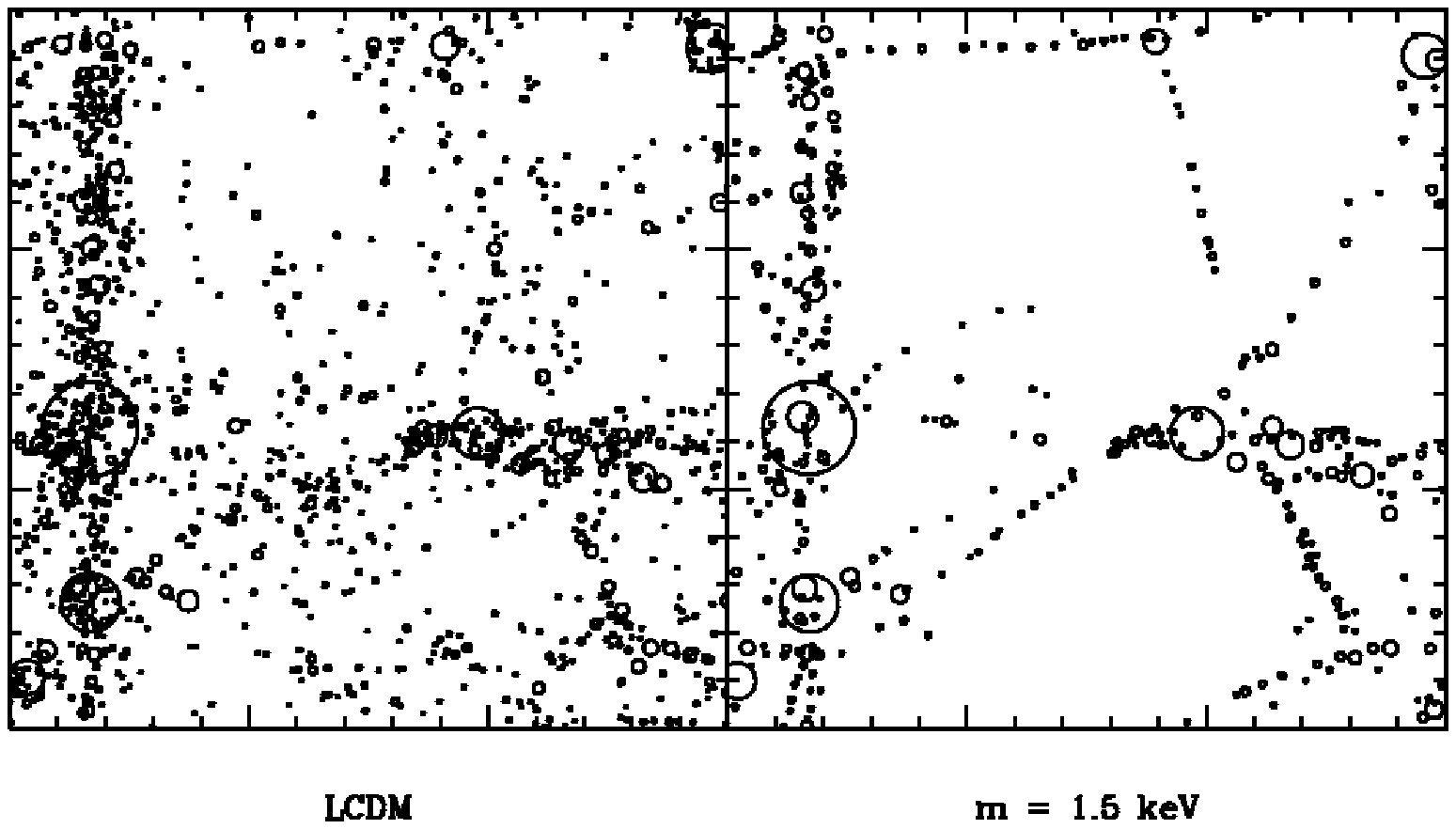}
\figcaption{Position of gravitationally bound halos at redshift $Z=0$
in a 3 $h^{-1}$Mpc box, for $\Lambda$CDM (left) and
$\Lambda$WDM with $m_X=1.5$ keV (right).  The radius of each
circle is $r_{200}$.  The smallest halo mass is
$2.7\times 10^7 h^{-1}M_{\odot}$. Finite box size effects are visible in
the orientation of the largest pancake, but it is clear that
the environment of small halos is distinct between the two scenarios.
\label{fig:1p5kmorph} }

\epsscale{0.8}
\plotone{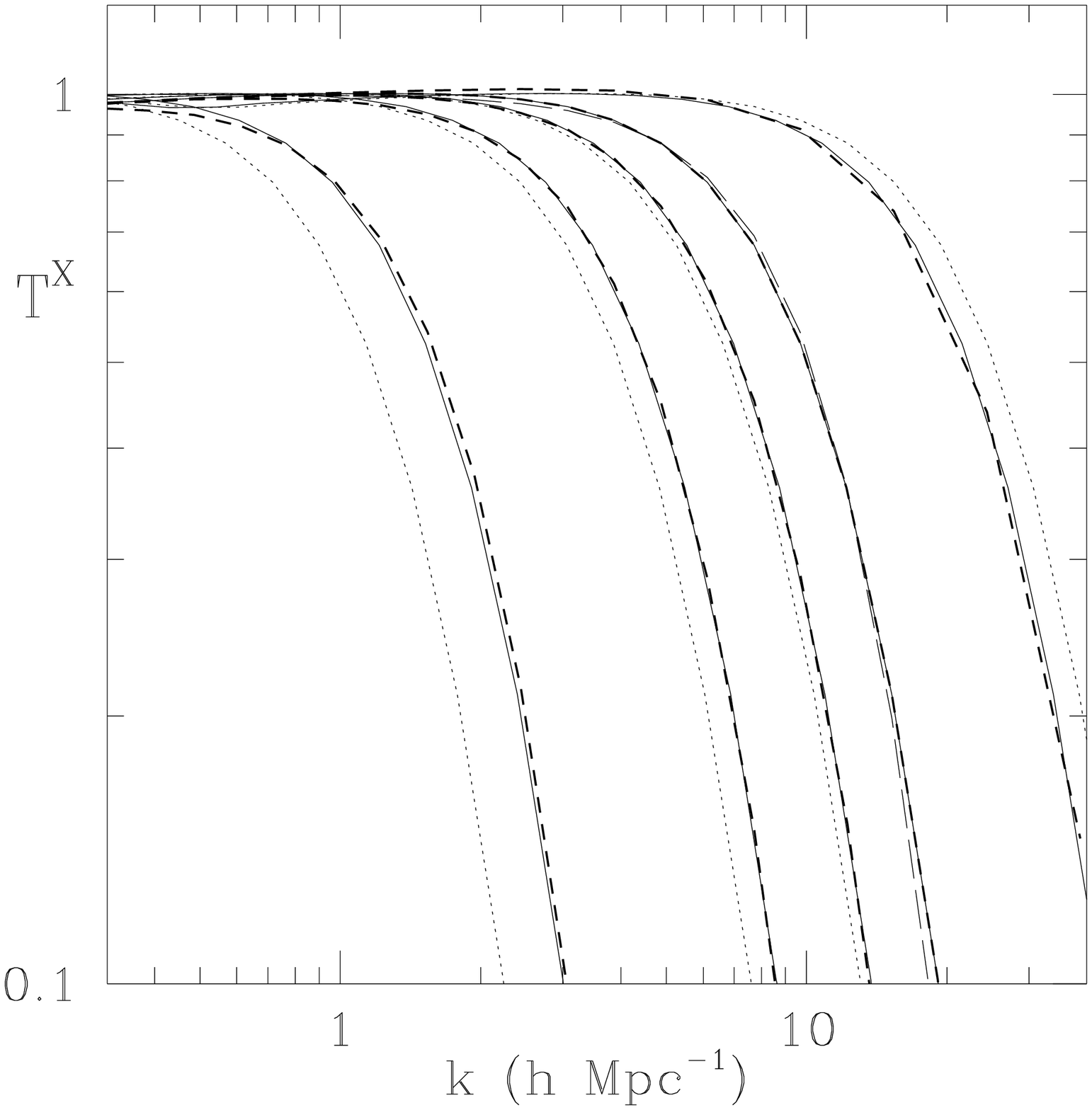}
\epsscale{1.0}
\figcaption{The transfer function  $T^X_k \equiv \sqrt{P^X_k/P_k^{CDM}}$
as computed with an accurate Boltzmann code \citep{Ma}, 
for $m_X$= 0.2 0.5 0.75, 1.0 and 2.0 keV, 
(bold dashed curves, left to right). The 1 keV curve is
then re-plotted after rescaling $k$ by $m_X^{1.15}$, for
each mass, as solid lines. The same  curve is also re-plotted
after the rescaling of $m_X^{4\over 3}$, as dotted lines,
demonstrating that this scaling is incorrect.
The fit we have adopted in equation (\ref{eq:txk}) is shown as
a long-dashed curve.
\label{fig:scaling} }


\end{document}